\documentclass[twocolumn]{aa}
\usepackage{graphicx}
\begin{document}
\title{ Ground-Based CCD Astrometry with Wide Field Imagers.
I.\thanks{Based on observations with the 2.2m MPI ESO telescope.}}

\subtitle{Observations just a few years apart allow decontamination of
          field objects from members in two Globular clusters. }

   \author{Jay Anderson\inst{1}, 
           Luigi R.\ Bedin\inst{2}, 
	   Giampaolo Piotto\inst{3}, 
	   Ramakant Singh Yadav\inst{3,4}, \and 
	   Andrea Bellini\inst{3}.
          }

   \offprints{L. R. Bedin}

   \institute{
	     Dept.\ of Physics and Astronomy, MS-108,
             Rice University, 6100 Main Street, Houston, TX-77005, USA\\
	     \email{jay@eeyore.rice.edu}
         \and
             European Southern Observatory, Garching,
	     Karl-Schwarzschild-Str.\ 2, D-85748, D, EU\\
             \email{lbedin@eso.org}
       	 \and
             Dipartimento\ di Astronomia, Universit\`a di Padova, 
	     Padova, I-35122, EU\\
              \email{piotto-bellini@pd.astro.it}
	 \and
	     Aryabhatta Research Institute of Observational Sciences (ARIES), 
	     Nainital, 263129, India.\\
	     \email{rkant@upso.ernet.in}
   }

    \date{Received 10 February 2006 / Accepted 15 April 2006}

   \abstract{ 
%%%
This paper is the  first of a series of papers in  which we will apply
the  methods  we have  developed  for  high-precision astrometry  (and
photometry) with the Hubble Space  Telescope to the case of wide-field
ground-based images.  In particular,  we adapt the software originally
developed for WFPC2 to ground-based, wide field images from the WFI at
the ESO 2.2m telescope.  In this paper, we describe in details the new
software, we  characterize the  WFI geometric distortion,  discuss the
adopted local transformation  approach for proper-motion measurements,
and  apply the  new technique  to two-epoch  archive data  of  the two
closest Galactic globular  clusters: NGC 6121 (M4) and  NGC 6397.  The
results of this  exercise are more than encouraging.   We find that we
can  achieve a  precision of  $\sim$7 mas  (in each  coordinate)  in a
single  exposure for  a well-exposed  star, which  allows a  very good
cluster-field  separation in both  M4, and  NGC~6397, with  a temporal
baseline of only 2.8, and 3.1 years, respectively.

%%%    
   } 
   \titlerunning{Wide Field CCD Ground-Based Astrometry. I. }
   \authorrunning{Anderson et al.}
   \maketitle
%________________________________________________________________

%%%%%%%%%%%%%%%%%%%%%%%%%%%%%%%%%%%%%%%%%%%%%%%%%%%%%%%%%%%%%%%%%
%
\section{Introduction}
\label{intr}
%
%%%%%%%%%%%%%%%%%%%%%%%%%%%%%%%%%%%%%%%%%%%%%%%%%%%%%%%%%%%%%%%%%

Recent investigations have shown that imaging from the cameras onboard
the {\em  Hubble Space Telescopes} ($HST$)  can provide high-precision
astrometry for  point-like sources  (Anderson \& King  2000, hereafter
AK2000).  There are several factors which make imaging astrometry much
more accurate from space than  from the ground.  First, the absence of
atmospheric  effects allows us  to obtain  diffraction-limited images,
with a point-spread  function (PSF) which is nearly  constant in time,
and  therefore  amenable  to  detailed  modeling.   Also,  space-based
observatories  are  free  of  differential-refraction  effects,  which
plague  ground-based images  not taken  at the  zenith.   Finally, the
weightless environment  means that telescope flexure does  not lead to
large changes in the distortion solution, which means we can model the
solution to much higher accuracy.

However, despite  all the benefits  of imaging astrometry  from space,
there are  some significant limitations  as well.  First, the  need to
download all  the data taken to the  ground puts a major  limit on how
much data  can be collected by  $HST$ per hour.  For  this reason, the
largest  detectors  are  4096$\times$4096,   and  almost  all  of  the
detectors are  undersampled in order  to get the maximum  sky coverage
for the  limited number of pixels.  Ground-based  telescopes suffer no
such  limitations.  They  can be  made up  of dozens  of CCDs  and can
collect  Terabytes  of  information  every  night.   Furthermore  each
exposure can  cover over 400$\times$ the  biggest $HST$ field-of-view.
In  addition,  the  fact  that  $HST$  is  undersampled  introduces  a
significant  complexity to the  data analysis.   Special care  must be
taken to  derive exquisitely accurate  PSFs (see AK2000), so  that the
positions  measured with them  will be  free from  bias.  Ground-based
detectors can afford to oversample the stellar image, so that sampling
will not be  a limitation or complication for  our accuracy.  Finally,
the fact  that time on $HST$  is scarce means  that it is hard  to get
space-based observations.   By contrast,  there are many  ground based
observatories. 

Even with  its sampling and field-of-view  limitations, the phenomenal
astrometric precision possible with  $HST$ has allowed us to undertake
projects that were simply impossible before, such as:
\begin{itemize}
 \item The geometrical determination  of the globular cluster distance
       scale by comparison of the internal proper motions, with radial
       velocity dispersion obtained from ground (Bedin et al.\ 2003a);
 \item The study  of  the low-mass  Main  Sequence (MS)  down to  the
       hydrogen burning limit  (King et al. 1998, 2005,  Bedin et al.\
       2001);
 \item The proper motions of the Galactic Globular clusters and nearby
       galaxies; (Bedin et al.\ 2003b, Milone et al.\ submitted);
 \item The Galactic dynamic measurements (Bedin et al. 2003b);
 \item The cluster rotation on the plane of the sky (Anderson \& King 2003a); 
 \item The double main sequence in Omega Centauri (Bedin et al. 2004);
 \item The anomalous white dwarf cooling sequence in the open cluster
  NGC 6791 (Bedin et al. 2005a).
\end{itemize}

Nonetheless, even  given the  clear advantages of  $HST$, there  are a
number  of cluster  studies  that are  better  suited to  ground-based
observations.   For  instance,  studying  the  outskirts  of  clusters
requires large  areal coverage but  does not require a  telescope with
the resolution of  $HST$.  Many aspects of cluster  evolution can only
be understood  by putting  together surveys done  in the  cluster core
with more extended surveys of  the outer regions. Therefore, large FOV
ground-based studies  are very much complementary to  the core studies
possible only  with $HST$.  
These  large-FOV  studies will  tend  to  probe  the clusters  in  the
outskirts  where  the  density  is  low  relative  to  the  field,  so
proper-motion cleaning  will play an essential role  in constructing a
pure cluster sample, as it has in many HST projects.

Our  interest  in  the   above  applications,  in  particular  in  the
proper-motion aspects,  has stimulated the effort to  transfer what we
have learned  by measuring high-accuracy positions on  $HST$ images to
wide field,  ground-based data.  Much attention has  been devoted over
the  years   to  software  that  can  extract   good  photometry  from
ground-based images (DAOPHOT, ROMAPHOT, etc), but thus far very little
attention has  been devoted to astrometry.   Photometry and astrometry
make  very different  demands  on PSF  analysis.  Photometry  concerns
itself  more with  sums  of  pixels, whereas  astrometry  keys off  of
differences between nearby  pixel values, but there is  no reason that
with  a good PSF  we cannot  measure both  good fluxes  and positions.
This paper is one step in that direction.

%--------------------------------------------------------------
%
% FIGX --- 
%
%--------------------------------------------------------------
%
   \begin{figure}
   \centering
   \includegraphics[width=9.cm]{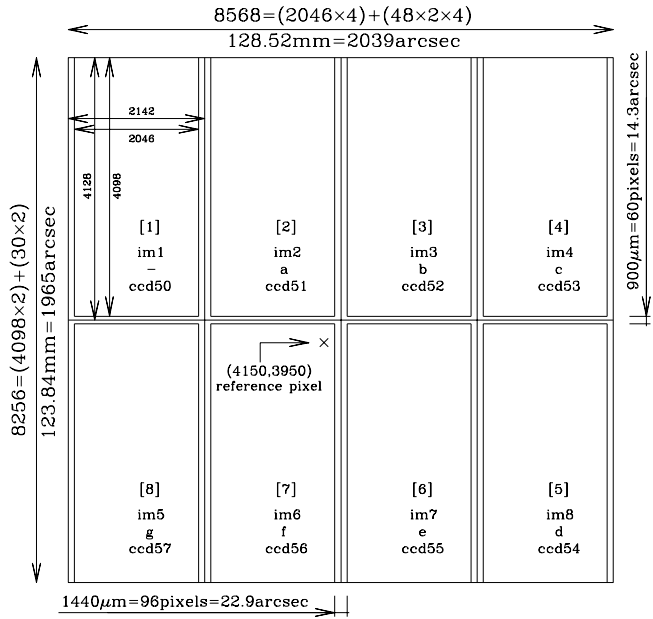}
      \caption{$WFI@2.2m$  layout. Dimension of  the chips,  gaps, and
		the  whole field  of  view, are  expressed in  pixels,
		linear units, and arcsec. Also each chip has different
		labels, in this paper we  will refer to each chip with
		the  numbers going  from [1]  to [8]  as shown  in the
		figures. }
         \label{wfi}
   \end{figure}
%_________________________________________________________________
%

%--------------------------
% FIGX --- 
%
%--------------------------------------------------------------
%
   \begin{figure*}
   \centering
   \includegraphics[width=16.cm]{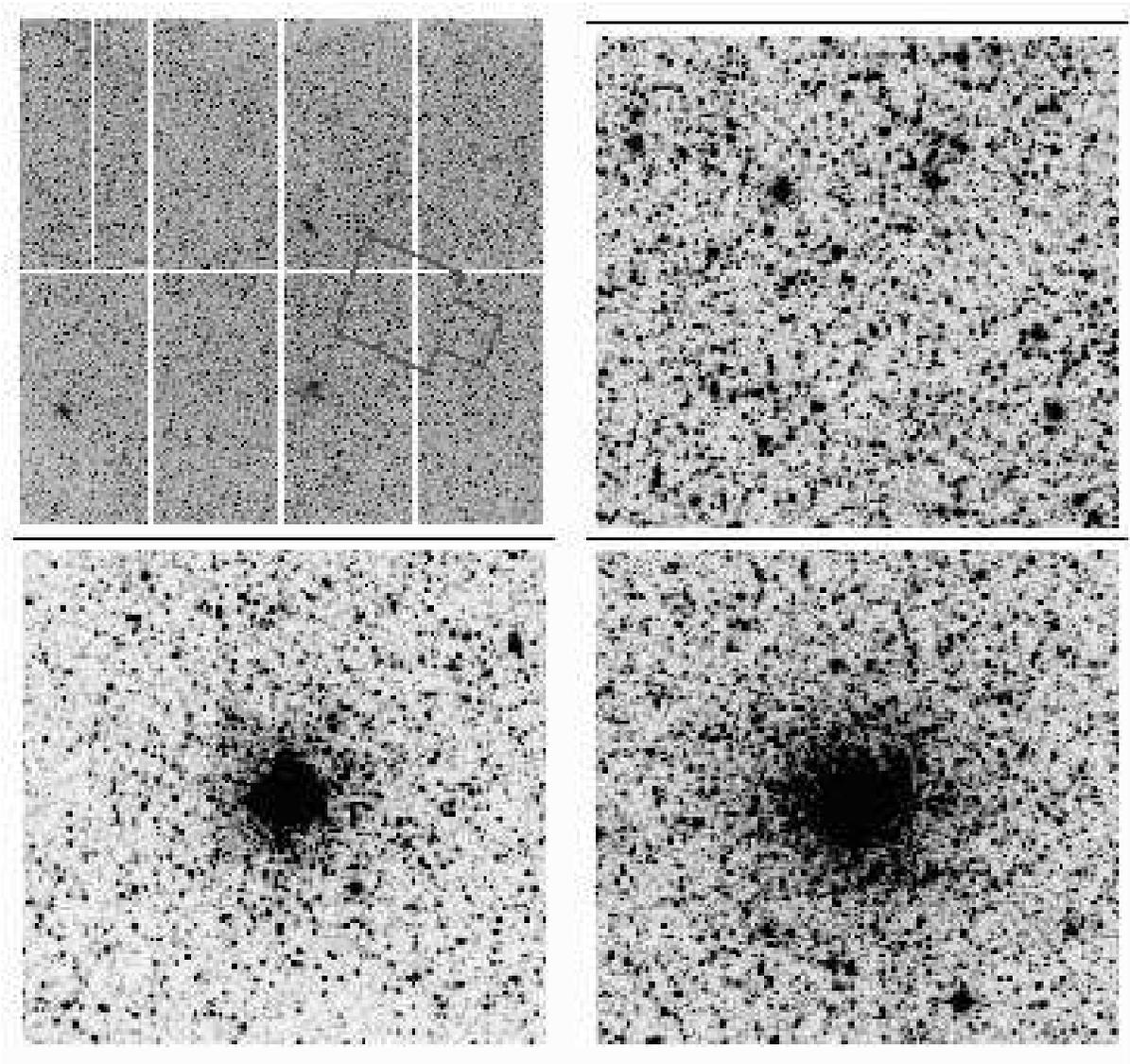}
      \caption{  
%
% FIGX --- 
%--------------------------
%\begin{figure*}[ht!]
%\begin{center}
%\begin{tabular}{cc}
%\fbox{\includegraphics[width=7cm]{f01.ps}}&\fbox{\includegraphics[width=7cm]{f02.ps}} \\
%\fbox{\includegraphics[width=7cm]{f03.ps}}&\fbox{\includegraphics[width=7cm]{f04.ps}} \\
%\end{tabular}
%\protect\caption[]{
%
{\em (top-left)} One of the images of Baade's window used to calibrate
the geometrical distortion  of $WFI@2.2m$. In these images  we see the
Galactic  globular  cluster  NGC~6528  on  chip[8],  and  NGC~6522  on
chip[6].  The dark spot on chip  [3] is just a very bright star.  Also
we over-plot the  footprint of a mosaic of  5 WFC/ACS fields available
from the archive, that we  used to cross check the astrometry obtained
with $WFI@2.2m$ (cfr.\ Sect.\ \ref{HSTcomp}).  This image gives a feel
for  the enormous  amount of  sky wide-field  imagers can  cover  in a
single exposure.
{\em (top-right)} A representative sub-set of the image, which show
the homogeneous distribution of stars in Baade's window.
{\em  (bottom-left)} Zoom-in  of ~1000$\times$1000  pixels  around the
globular cluster NGC~6528.
{\em (bottom-right)} Zoom-in of the globular cluster NGC~6522.
}
%\label{baadeW}
%\end{center}
%\end{figure*}

\label{baadeW}
\end{figure*}
%_________________________________________________________________
%

Over the last  few years, several Wide Field  Imagers (WFIs) have come
on-line at  large ground-based telescopes  (MPI-ESO 2.2m, AAT  4m, CFH
4m),  and  their  number  and  their  field-of-view  are  continuously
increasing (LBT 2$\times$8m, VST  2.5m, UKIRT 3.8m, VISTA 4m, etc...).
These WFIs allow us to map  completely any open or globular cluster in
our Galaxy and  their tidal tails, and to  get accurate photometry for
enormous numbers of stars.

One of the most promising (yet still largely unexplored) opportunities
presented   by  wide-field   images  involves   astrometry.   Accurate
astrometry over wide fields is important for a variety of reasons.  To
be sure,  an accuracy of 0.2  arcsec or better is  usually required to
position point-like sources in the increasing number of multi-slit and
multi-fiber spectroscopic facilities.
But the most promising astrometric  applications lie in the ability to
measure proper-motions for a large number of stars.  In principle, the
ground-based  WFIs  should  allow  astrometric  measurements  with  an
accuracy far better than the nominal $0.2$ arcsec.  As we will show in
Sect.\ \ref{APPLICATION}, with a baseline  of just a few years, images
collected with  modern WFIs can  provide proper motions  more accurate
than  those obtainable  with old  plates  with a  baseline of  several
decades.  (Note, though, that  these plates will still remain valuable
for long-term non-linear astrometry,  such as the determination of the
orbit  of long-period  visual binaries,  and of  course  for long-term
variation in the light curves).

In this paper, we apply what we have learned from $HST$ to the case of
one particular  ground-based wide-field imager:\ the WFI  at the focus
of the 2.2m ESO/MPI telescope,  located at La Silla (hereafter we will
refer to it as  $WFI@2.2m$).  The WFI camera is made up  of 8 chips of
2142$\times$4128  pixels   each  disposed  as   illustrated  in  Fig.\
\ref{wfi}, with a pixel-scale of 238 mas/pixel.

The reason  for choosing this  detector is that the  $WFI@2.2m$ camera
was  one  of the  first  wide field  cameras  to  become available  to
astronomers.  It began  its operation in 1999, and  today there are in
the public archive many  multiple-epoch images of star cluster fields,
with baselines up to 6 years.

There  are  clearly  some  things  that  only  $HST$  can  do,  namely
astrometry and photometry of extremely faint stars or stars in crowded
regions,  where there  truly  is no  substitute  for high  resolution.
Nevertheless,  we  show  here  that  many  scientifically  interesting
projects can now be carried out with ground-based imagers, such as the
$WFI@2.2m$ (discussed here) or the OMEGACAM (coming on-line in 2006).

In this paper, we will go  through the steps that are necessary to get
good  astrometry with  wide-field  detectors. In  Section  2, we  will
describe  the database  used for  this work.   Section  \ref{PSF} will
describe the method used to construct accurate PSFs. Section \ref{XYM}
will  give  details  on  the  fitting  procedure,  including  neighbor
subtraction.   In  Section \ref{GC}  we  will  discuss the  distortion
correction, and  its stability  over time. Section  \ref{HSTcomp} will
compare  the  astrometry  obtained  with  $WFI@2.2m$  with  astrometry
obtained  from  $HST$ archive  images  of  the  same region.   Section
\ref{LOCAL}  will  describe  the  local-transformation  approach  that
allows  us to  minimize the  effects  of residuals  in the  distortion
corrections. In section \ref{APPLICATION}  we will apply the method to
the case  of the  two closest globular  clusters, namely  NGC~6397 and
NGC~6121 (M4).
In Sect.\ \ref{ATM} we briefly discuss atmospheric effects.
Finally, in  section \ref{CON} we  summarize our results,  and briefly
discuss possible interesting projects for the future.

%%%%%%%%%%%%%%%%%%%%%%%%%%%%%%%%%%%%%%%%%%%%%%%%%%%%%%%%%%%%%%%%%
%
\section{Observations}
\label{OBS}
%
%%%%%%%%%%%%%%%%%%%%%%%%%%%%%%%%%%%%%%%%%%%%%%%%%%%%%%%%%%%%%%%%%

In this work,  we use images for four  different fields collected with
the $WFI@2.2m$; details of the data can be found in Tab.\ \ref{obs}.

The first field is located in Baade's Window (see Fig.\ \ref{baadeW}).
Although the field contains  two small globular clusters (NGC~6522 and
NGC~6528), most  of the  field has a  smooth, uniform  distribution of
Galactic bulge  stars.  The stars of interest  (the $\sim$2 magnitudes
below saturation in a 60s  $V$ exposure), are typically separated by a
few arcseconds, so that there are  many in each field, but they are in
general well  enough separated to  allow accurate positions.   We took
images of this field with a  range of offsets so that we could measure
the distortion in the detector  and evaluate its stability (cfr Sect.\
\ref{GC}).

The second  field is centered  on the open cluster  NGC~2477.  Several
long exposures  are taken  in almost identical  conditions (comparable
seeing, no large offsets,  identical exposure times). For this reason,
and  thanks to  the ideal  stellar density,  it has  been  possible to
estimate directly  the internal photometric and  astrometric errors of
our method from analysis of the residuals (cfr.\ Sect.\ \ref{dein}).

The third field used in this work is centered on NGC~6397.  The images
were taken  at two different epochs  separated by 3.1 yrs. We will use
the two epochs to derive proper motions (Sect.\ \ref{NGC6397}).

The fourth field covers M4.  Images were taken at two different epochs
separated by 2.8 yrs.  Also for this object we will use the two epochs
to  derive proper  motions and  distinguish cluster  stars  from field
stars (Sect.\ \ref{M4}).

In  anticipation  of  the  need  to reduce  the  enormous  archive  of
$WFI@2.2m$   data  in   an  automated   way,  we   developed  software
specifically for  this instrument, though  the software can  be easily
adapted  to other  CCD  mosaics, including  OMEGACAM.  One  particular
effort we  make to  deal with the  huge images involved  in wide-field
surveys  is that  we take  care to  do all  stages of  reduction  in a
short-integer format.   This improves  read-in and read-out  time, and
helps  enormously with diskspace  considerations (256Mb  compared with
$\sim$60Mb once gzipped).

%__________________________________________________ 
% 

\begin{table} 
\caption{Description of the data-set used for this work.}  
\centering 
\label{obs}     
\begin{tabular}{cccc} 
\hline\hline    
filter & EXP-TIME & seeing & airmass\\ 
\hline\hline          
%------------------------------------------------------------------------------------------ 
& & & \\
\hline
\multicolumn{4}{c}{ {\bf Bulge --- Baade window} } \\
\hline    
& & & \\
\multicolumn{4}{c}{calibration data, June 6, 2003}\\
& & & \\
 $U$  &   3$\times$350s;    &  $\sim 1''.3$    & 1.18-1.23 \\
 $V$  &  30$\times$60s;     &  $0''.7$-$1''.6$ & 1.00-1.13 \\
& & & \\
%------------------------------------------------------------------------------------------
& & & \\
\hline
\multicolumn{4}{c}{ {\bf NGC~2477} } \\
\hline    
& & & \\
\multicolumn{4}{c}{test data, January 20, 1999}\\
& & & \\
 $I_{\#853}$  &  6$\times$900s;     &  $0''.9$-$1''.1$ & 1.05-1.33 \\
& & & \\
%------------------------------------------------------------------------------------------
\hline          
\multicolumn{4}{c}{ {\bf NGC~6121 (M4)} }\\
\hline          
& & & \\
\multicolumn{4}{c}{Epoch I: August 17-18, 1999}\\
& & & \\
 $B$  & 3$\times$180s;                     & $\sim 1''.3$ & 1.04 \\
 $V$  & 3$\times$180s;                     & $\sim 1''.5$ & 1.20 \\
& & & \\
\multicolumn{4}{c}{Epoch II: June 21, 2002}\\
& & & \\
 $B$  & 3$\times$5s;    1$\times$100s;    & $\sim 1''.4$ & 1.20 \\
 $V$  & 3$\times$10s;   6$\times$90s;     & $\sim 1''.1$ & 1.10 \\
& & & \\
%------------------------------------------------------------------------------------------
\hline          
\multicolumn{4}{c}{ {\bf NGC~6397}}\\
\hline          
& & & \\
\multicolumn{4}{c}{Epoch I: May 14, 1999}\\
& & & \\
 $B$  & 1$\times$20s;    2$\times$240s;    & $\sim 1''.1$ & 1.10 \\
 $V$  & 1$\times$20s;    2$\times$240s;    & $\sim 1''.2$ & 1.08 \\
 $I$  & 1$\times$20s;    2$\times$240s;    & $\sim 1''.0$ & 1.09 \\
& & & \\		                    
\multicolumn{4}{c}{Epoch II: June 18, 2002}\\
& & & \\
 $U$  & 3$\times$35s;    3$\times$240s;    & $\sim 1''.4$ & 1.08 \\
 $B$  & 3$\times$5s;     3$\times$90s;     & $\sim 1''.2$ & 1.10 \\
 $V$  & 3$\times$3s;     8$\times$70s;     & $\sim 1''.3$ & 1.11 \\
 $I$  & 1$\times$3s;     2$\times$3s;      & $\sim 1''.1$ & 1.21 \\
      & 1$\times$49s;    3$\times$50s;     & $\sim 1''.1$ & 1.21 \\
& & & \\
%------------------------------------------------------------------------------------------
\hline                                   %inserts single line 
\end{tabular} 
\end{table} 
%__________________________________________________ 
% 

%
%_________________________________________________________________
%

%%%%%%%%%%%%%%%%%%%%%%%%%%%%%%%%%%%%%%%%%%%%%%%%%%%%%%%%%%%%%%%%%
%
\section{Derivation of the PSF}
\label{PSF}
%
%%%%%%%%%%%%%%%%%%%%%%%%%%%%%%%%%%%%%%%%%%%%%%%%%%%%%%%%%%%%%%%%%

Anderson  \& King (2000)  developed a  method to  obtain high-accuracy
astrometry on  under-sampled WFPC2 images.   A careful removal  of all
the  sources  of  systematic  errors,  such as  biases  introduced  by
under-sampling,  chip-manufacturing  defects  (see  Anderson  \&  King
1999), and the need for  an accurate correction for distortion, led us
to  arrive at  what is  more or  less {\em  the state  of the  art} in
imaging astrometry from space.

We found in our treatment of $HST$ images that astrometry is even more
sensitive to  the PSF  model than photometry  is.  This is  because to
first order, photometric procedures  don't really care {\it where} the
flux  is, so  long as  it is  included within  the fitting  radius (or
aperture).   Astrometric procedures,  on the  other hand,  key  off of
exactly how  the flux is  distributed among the pixels.   We therefore
require  an accurate PSF  to compare  the observed  pixel distribution
with the model, in order to extract a position.

Even though the WFI detectors  are not undersampled, our positions are
still  critically  dependent  on  the  accuracy  of  our  PSF  models.
Thankfully, it is much easier  to derive PSFs from well-sampled images
than  from undersampled  ones, so  that much  of the  careful  work in
AK2000 is not required in  this regime.  In particular, it is possible
to derive a  PSF from a single image, without  reference to a dithered
set.   This is  good news,  since thanks  to seeing  fluctuations, the
ground-based  PSF cannot  be presumed  to be  stable from  exposure to
exposure.

%%%%%%%%%%%%
\subsection{The empirical PSF model}
%%%%%%%%%%%%

A PSF model simply tells us what fraction of a star's flux should fall
in a pixel located at a  given offset from the star's center.  The PSF
is therefore a two-dimensional function $\psi(\Delta x,\Delta y)$ that
returns, for a given $(\Delta x,\Delta y)$, the fraction of light that
would fall in a pixel at that offset.

Unlike DAOPHOT  (Stetson 1987) and other photometry  routines, our PSF
model does not have an  analytical-function as a backbone.  Rather, we
represent  the PSF  entirely by  an empirical  grid, a  simple look-up
table.  The  entire PSF is  represented by an array  of 201$\times$201
grid points.   This PSF grid  is super-sampled by  a factor of  4 with
respect  to the image  pixels, so  that the  PSF model  goes out  to a
radius of about 25 pixels.   The central grid point at (101,101) tells
us what fraction of a star's flux would land in the central pixel of a
star that is centered on a  pixel.  The other grid points tell us what
flux would  fall in pixels at  an array of  quarter-pixel offsets from
the star's center.  Of course, stars can land anywhere within a pixel,
and not just at quarter-pixel grid points, so we use a bi-cubic spline
to interpolate the value of the PSF in between the grid points.

%%%%%%%%%%%%
\subsection{Finding the PSF by iteration}
%%%%%%%%%%%%
%
Following the  above definition, we can  predict the value  of a given
pixel $(i,j)$  in the vicinity of a  star if we know  the star's total
flux $z_*$, it's position $(x_*,y_*)$, and the sky background $s_*$:
$$
   P_{ij} = z_* \cdot \psi(i\!-\!x_*,j\!-\!y_*) + s_*,
$$
For each star, we have an array  of pixels that we can fit in order to
solve for the triplet of parameters: $x_*$, $y_*$, and $z_*$.  The sky
$s_*$ is usually constrained by a more remote annulus.

If we have a  set of positions and fluxes for a  star, we can turn the
above equation around to solve for the PSF:
$$
   \psi(\Delta x,\Delta y) = (P_{ij}-s_*)/z_*.
$$
This equation  means that each pixel  in the star's image  gives us an
estimate  of the  2-dimensional  PSF function  at  one point---at  the
location $(\Delta x,\Delta y) = (i-x_*,j-y_*)$.  We construct a single
general PSF model by combining  the array of samplings from many, many
stars.

The derivation of  an accurate PSF is clearly  an iterative procedure.
Without  a good  PSF,  we  cannot derive  good  positions and  fluxes.
Similarly,  without good  positions and  fluxes, we  cannot  derive an
accurate PSF.  Thus,  our procedure iterates in order  to improve both
the stellar parameters and the PSF model.

We start with simple centroid positions and aperture-based fluxes.  It
does not  take many iterations to  arrive at good models  for both the
PSF and the stellar parameters.  Our iterative procedure here is quite
similar to  that of AK2000, except  that we do not  require the second
stage  of their  three-stage  iteration.  Since  our  images are  well
sampled,  we  do not  need  to  incorporate  images taken  from  other
ditherings to  remove the star-position/PSF-shape  degeneracy inherent
in  undersampled detectors.  This  simplifies our  reduction procedure
significantly, since we can now operate on one exposure at a time.

%%%%%%%%%%%%
\subsection{Constraints on the PSF}
%%%%%%%%%%%%
%
We  chose  to use  a  grid-based  model for  the  PSF  because of  its
flexibility.  A simple grid makes it  very easy to adjust the shape of
the PSF in exactly the place the  data say that it may need to change.
Sometimes, however,  such a grid  can have {\it too}  much flexibility
and can bend in unphysical  ways.  Thus, we impose some constraints to
ensure a reasonable PSF.

The first  constraint we apply  is smoothness.  Since our  detector is
well-sampled, the  PSF should  not change too  much from  gridpoint to
gridpoint.   We enforce  this by  smoothing the  PSF with  a quadratic
smoothing  kernel (again, see  AK2000).  The  quadratic kernel  fits a
quadratic  to  the  gridpoints  within  the kernel  centered  on  that
gridpoint  (5$\times$5 gridpoints,  7$\times$7 gridpoints,  etc), then
replaces the  central value  with the value  of the quadratic  at that
point.  We experimented and  adopted the largest smoothing kernel that
was consistent with the star  images.  (If too much smoothing is done,
stars have large residuals at their centers.)

The second constraint  we apply is that the PSF has  to be centered on
the grid.  To enforce this,  we fit the central 11$\times$11 pixels of
the PSF with  a paraboloid, to estimate the  apparent center.  If this
center  is  not at  the  center  of the  grid,  then  we use  bi-cubic
interpolation  to  re-sample  the  PSF  at  the  locations  where  the
gridpoints should  be and replace  the PSF with the  properly centered
model.

The  final  constraint  we  apply  is  normalization.   It  is  nearly
impossible to  measure all  the flux in  the PSF.  The  finite dynamic
range of detectors  means that stars which are  not saturated in their
cores are lost in the sky  noise beyond about 8 pixels.  The saturated
stars can be  seen well beyond this, but it is  hard to determine what
fraction of their flux we  are seeing, since their central regions are
unusable.  Our  routine does derive a  PSF out to 25  pixels using the
saturated stars, but it is most  accurate within 8 pixels, where it is
derived  from  well-measured   bright  unsaturated  stars.   Thus,  we
normalized the  PSF to  have a  volume of unity  within 6  pixels (1.5
arcseconds), so  that our normalization  would not be affected  by the
uncertainties  related to saturation.   The $WFI@2.2m$  PSFs typically
have 15 percent of their flux beyond this radius, so determining total
fluxes will  require a simple  aperture correction in  the calibration
process.

%%%%%%%%%%%%
\subsection{Variability of the PSF with chip position}
%%%%%%%%%%%%
%
Our PSF-modeling procedures allow us to evaluate directly how well the
PSF  fits stars.  We  initially derived  a single  PSF for  the entire
8-chip  detector,  but  we  soon  found  that  there  were  systematic
residuals in the stellar profiles,  indicating that the PSF was indeed
quite different from one chip to another.  Then we solved for a single
PSF for  each chip.  Again, the  residuals indicated that  the PSF was
changing shape  from one part of  the chip to the  other.  This change
was in  fact quite significant: the  fraction of flux in  the core was
seen to vary by $\pm$10\%.

\begin{figure}
   \centering
   \includegraphics[width=8.9cm]{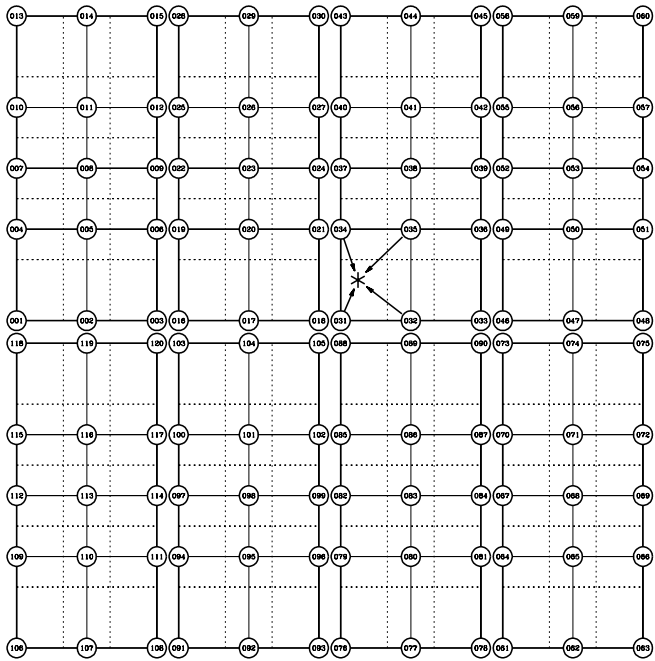}
      \caption{
This  figure shows  the locations  of the  120 fiducial  PSFs  and the
process of  interpolation to find a  PSF at a particular  point on the
chip (marked with an \textbf{\Large $\ast$}).  The dotted lines denote
the region of the image used to solve for each PSF.
      }
         \label{interp}
   \end{figure}

In the end, we  determined that by solving for an array  of 15 PSFs in
each  2048$\times$4096-pixel chip  (3  across and  5  high), we  could
capture almost all of the PSF's spatial variability.  Our model, then,
will  feature an independent  PSF at  each of  the locations  in Fig.\
\ref{interp}. \\

%
% FIGX --- 
%______________________________________________________________
%
%_________________________________________________________________
%

\begin{figure}[ht!]
   \centering
   \includegraphics[width=8.5cm]{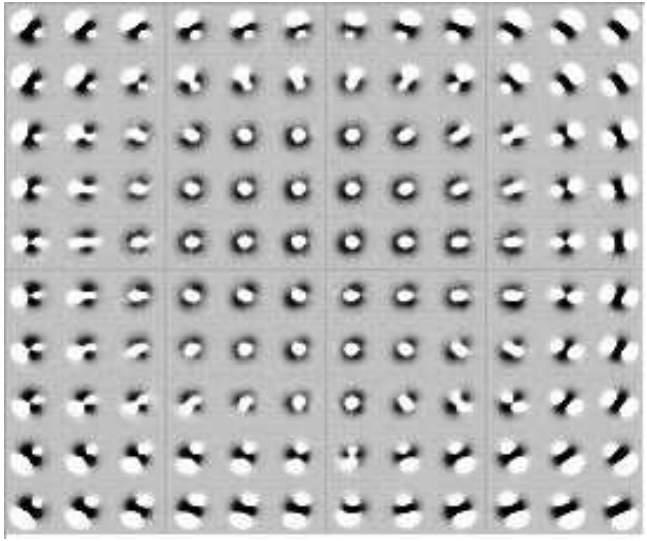}
      \caption{Difference between the local PSF and the average PSF
               over the entire field. 
               White means more flux in the local PSF than in 
	       the average PSF at that location.}
         \label{psf}
   \end{figure}

To construct a model for the  PSF in between these fiducial points, we
will use simple linear interpolation, in a manner similar to AK2000.
To highlight how the PSF  changes shape with location in the detector,
in Fig.\  \ref{psf} we show  the difference of the  120 representative
fiducial PSFs  and the  average PSF across  the entire field  of view.
The PSF  comes from  one of the  $U$-band Baade's window  images.  The
PSFs  are  conveniently  displayed  in  order  to  map  their  spatial
distribution  on  the  detector.   Notice  how  the  variations  among
contiguous PSFs are smooth.  The  PSFs tend to elongate (mainly due to
coma aberration),  with the elongation increasing radially  as we move
out from the principal optical axis (near the center of the detector).

It  is obvious  from  Fig.\ \ref{psf}  that  it would  be possible  to
reparametrize the PSF and reduce the number of degrees of freedom used
by taking  advantage of  the clear radial  behavior.  But our  aim has
been to  minimize the amount  of human intervention required,  even at
the expense of over-parametrizing  the PSFs. (This approach applies to
our distortion solution, too).  Even though we may use more parameters
than necessary, each of our  PSFs is still greatly overconstrained, so
there  is no  real  advantage  in seeking  a  slightly more  efficient
parametrization.  With our  very general parametrization, our routines
can operate with a minimum of human oversight.

%%%%%%%%%%%%
\subsection{Choosing the stars used to model the PSF}
%%%%%%%%%%%%
%
In order  to tell us  something about the  PSF, a star must  have good
signal to noise  in both in the  core and beyond the core,  so that we
can determine from its pixels how the flux is distributed.  Good stars
for  the PSF  must  therefore have  a  minimum of  $5000$ DN  (digital
numbers) above sky  in their  central  3$\times$3 pixels  and have  no
nearby neighbors.  We like to have  at least 50 such stars for each of
the  fiducial   PSFs  we  are  solving   for,  so  that   we  have  an
over-determined  problem and  can  iteratively reject  stars that  are
compromised by nearby neighbors, cosmic rays, or detector defects.

The  PSF-finding program  is designed  to require  minimal interaction
with the user.  The user supplies the program with some simple finding
criteria (minimum  flux, minimum separation from  brighter stars), and
the program searches  the image to find stars  that meet the criteria.
The  program  then reports  how  many  stars  are available  for  each
PSF-region.  If there are more than  150 good stars in a single region
(see Fig.\ \ref{interp}), then the routine chooses the 150 best stars,
based on brightness and isolation from neighbors. \\
%

% FIGX --- 
%______________________________________________________________
%
%_________________________________________________________________
%

In  sparse fields (or  equivalently, short  exposures), there  are not
always enough  bright stars to use  in PSF construction,  and we often
must make a compromise between the number of stars available to define
the PSF  and the  level of PSF  variability we can  practically model.
For  this  reason,   the  program  has  been  set   up  to  allow  the
determination       of:\      (3$\times$5)$\times$8($=$120      PSFs),
(3$\times$3)$\times$8($=$72  PSFs), (2$\times$3)$\times$8($=$48 PSFs),
(2$\times$2)$\times$8($=$24 PSFs), or (1)$\times$8($=$8 PSFs) to cover
the detector.  Based on the number of bright stars available, the user
must determine how finely to model the PSF's spatial variability.

%%%%%%%%%%%%
\subsection{PSF storage}
%%%%%%%%%%%%
%
Once an array  of PSFs has been constructed for an  image, we save the
array of PSF gridpoints in a simple fits image, which can be inspected
easily by  eye. The image shown  in Fig.\ \ref{psf}  is the difference
between one such image and the average PSF for the entire field. 

%%%%%%%%%%%%
\subsection{Planning for the future}
%%%%%%%%%%%%
%
This routine has been designed with a lot of flexibility, so that when
larger wide-field arrays  come on line, it will be  easy to expand the
number of chips or  the number of PSFs per chips to  deal with the new
images in an automated way.  The  3$\times$5 array of PSFs allow us to
deal with  the fact  that most chips  are rectangular and  not square.
Simple quadratic  variation would treat  the two axes  differently and
the PSF quality would suffer.

%%%%%%%%%%%%%%%%%%%%%%%%%%%%%%%%%%%%%%%%%%%%%%%%%%%%%%%%%%%%%%%%%
%
\section{Fitting Star Positions and Flux}
\label{XYM}
%
%%%%%%%%%%%%%%%%%%%%%%%%%%%%%%%%%%%%%%%%%%%%%%%%%%%%%%%%%%%%%%%%%

% FIGX --- 
%--------------------------
\begin{figure*}[ht!]
\begin{center}
\includegraphics[width=16.5cm]{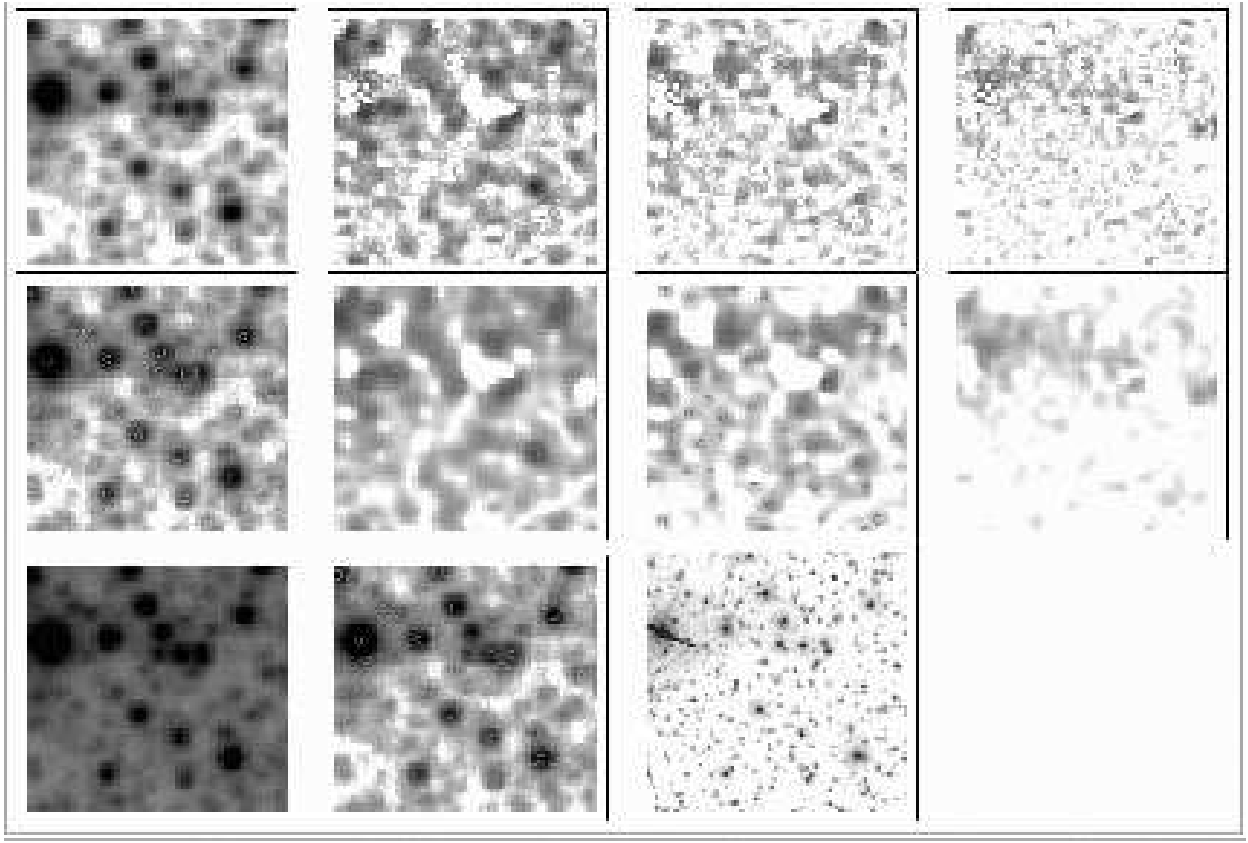}
% 
%\begin{tabular}{cccc}
%%
%\fbox{\includegraphics[width=3.8cm]{f07.ps}}&\fbox{\includegraphics[width=3.8cm]{f08.ps}} &
%\fbox{\includegraphics[width=3.8cm]{f09.ps}}&\fbox{\includegraphics[width=3.8cm]{f10.ps}} \\
%%
%\fbox{\includegraphics[width=3.8cm]{f11.ps}}&\fbox{\includegraphics[width=3.8cm]{f12.ps}} &
%\fbox{\includegraphics[width=3.8cm]{f13.ps}}&\fbox{\includegraphics[width=3.8cm]{f14.ps}} \\
%%
%\fbox{\includegraphics[width=3.8cm]{f15.ps}}& 
%\fbox{\includegraphics[width=3.8cm]{f16.ps}}&\fbox{\includegraphics[width=3.8cm]{f17.ps}}& \\
%%
%%\end{tabular}
\caption[]{
An image  subset of 70$\times$70  pixels ($\sim 17'' \times  17''$) is
shown through the various steps of star fitting.
On  the top,  the  sky-subtracted  image (corrected  for  CRs and  bad
pixels) through  the iteration 1, 2, 3,  and 7.  At each  step the new
difference  image (between  the  corrected and  the  model images)  is
calculated using the new  detected objects, and the improved positions
and fluxes of sources detected in the previous iteration.
The star finding  at each step is done  on the corresponding convolved
star-subtracted images (shown in middle line of figures).
For comparison,  in the bottom are  shown, the raw  image, the finding
chart of the final list of  detected objects, and the same area imaged
with WFC/ACS $HST$ (data set name {\sf j8kce1atq\_drz}).
}
\label{fitting}
\end{center}
\end{figure*}
%------------------------

Once an  array of PSF models has  been generated for an  image, we can
use it  to measure all the stars  in the image.  We  designed a simple
iterative  procedure that  seems to  work quite  well for  images with
sparse to moderate  crowding.  Once again, the routine  is designed to
have  minimal input  from  users.   The user  simply  inputs what  the
faintest findable star  should be (above sky) and how  close it can be
to brighter  neighbors, and the  program finds and measures  all stars
that  fit these  criteria.  Our  goal is  not to  give  a line-by-line
account here of what the program  does, rather we will simply give the
general procedure.

Fitting   overlapping  stellar  profiles   is  clearly   an  iterative
procedure.  The routine keeps in memory four images:
(1) the raw image, sky-subtracted, and corrected for cosmic rays (CRs) 
    and bad pixels/columns, 
(2) the model image, which has a properly scaled PSF for each found star, 
(3) the difference image, and 
(4) the convolved image.

Our  first iteration  begins by  finding all  the saturated  stars and
measuring fluxes for  them.  It also identifies all  peaks that are 10
times  brighter than  the threshold  and  10 pixels  farther than  any
brighter source,
and adds  them to  the star  list.  It measures  fluxes for  all these
stars using  the PSF,  then subtracts the  unsaturated stars  from the
image.  It  does not  subtract the saturated  stars, since the  PSF is
generally  not reliable  out in  the wings,  and that  would  make the
subtracted  image less useful  than the  original image.   The program
then takes this subtracted image  and convolves it with the PSF.  This
allows  the signal  from fainter  stars to  be optimally  enhanced for
finding  them.  We  also  generate  a model  image  which contains  an
estimate of the flux from all the stars, saturated and unsaturated.

The second iteration then finds stars in the convolved image, lowering
the threshold to  5$\times$ the minimum and insisting  again that they
be  isolated  from other  unfound  stars.   Any  stars found  in  this
iteration must be  at least $\sim$15\% brighter than  the model image,
so  that we  can  be  sure of  their  authenticity.  This  requirement
prevents us from  finding very faint stars next  to very bright stars,
but this is  not a severe limitation, as we could  not find such stars
reliably anyway.  The benefit of the requirement is that we do not end
up identifying a lot of undocumented PSF features as stars.

After  this second wave  of finding,  we re-solve  for all  the stars,
using for  each star an image  that has all  its neighbors subtracted.
This way, the  fits for two nearby stars can  quickly converge upon an
accurate  position and  flux for  each.  We  repeat  these iterations,
lowering the  threshold, and  incorporating fainter and  fainter stars
with each iteration.

In  Fig.\ \ref{fitting}  we  show  the various  steps  of the  fitting
procedure.  In practice,  we often have to run the  routine once or so
to determine how faint the final  threshold should be in order to find
and measure the faintest believable stars.

%%%%%%%%%%%
\subsection{Direct estimate of internal errors}
\label{dein}
%%%%%%%%%%%

We used multiple observations of the Galactic open cluster NGC~2477 (6
images of 900s in filter $I_{\#853}$)\footnote{
Note that  \#853 really  designates a number  ID, and not  a wavelength
(see                             also                             {\sf
http://www.ls.eso.org/\-lasilla/\-sciops/\-2p2/\-E2p2M/\-WFI/\-filters/}).
} 
to provide a direct estimate of the internal errors.  We could not use
the Baade's window data-set because of the large offsets (residuals in
the geometrical distortion would mask our internal errors).

\begin{figure}[!ht]
   \centering
   \includegraphics[width=9.cm]{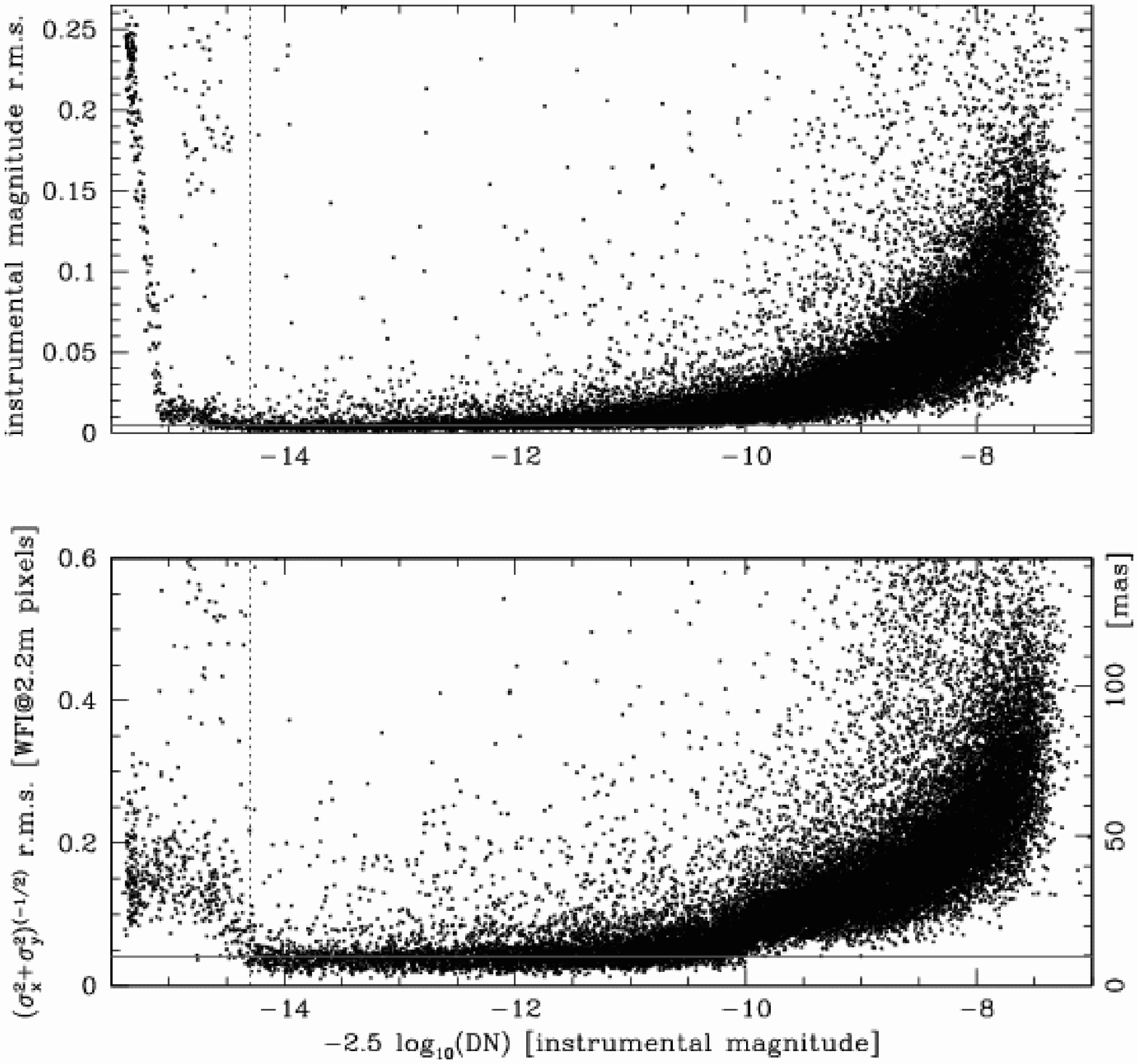}
      \caption{
%	%
	{\em  (Top  Panel):}  Photometric  errors per  exposure  as  a
        function of the instrumental magnitude defined as $-2.5 \times
        \log{\rm DN}$  (DN $=$ Digital  Numbers), obtained from  the 6
        archive images in filter $I_{\#853}$, with small off-sets, and
        identical  exposure times.   Each  star has  been measured  at
        least 6  times. In this data  set saturation sets  in at $\sim
        -14.3$,  and  is  indicated  by  a vertical  dotted  line.   A
        horizontal  line  shows the  average  r.m.s. for  well-exposed
        stars (0.005 mag).
%	%
	{\em  (Bottom  Panel):} For  the  same  data set,  astrometric
        r.m.s.  both in $WFI@2.2m$ pixels and in mas. Again the dotted
        vertical  line  marks  the  beginning  of  saturation,  and  a
        horizontal  line  shows the  average  astrometric r.m.s.   for
        well-exposed stars  0.04 pixels,  i.e. $\sim 10$  mas (meaning
        $\sim6.7$ mas  for each  coordinate).  Note how  positions are
        well  defined also  for moderately  saturated  stars ($\sim$35
        mas).
%	%
      }
         \label{rms1}
   \end{figure}

In  the top  panel of  Fig.\  \ref{rms1} we  show the  r.m.s.\ of  the
photometry as  function of  the instrumental magnitude.   A horizontal
line shows  that for well-exposed  stars, we attain  a single-exposure
internal precision of 0.005 magnitude.
In the bottom panel, we show  the behavior of the r.m.s.\ in position,
taken as the  sum in quadrature of the r.m.s.\ along  the $x$, and $y$
axes of  the detectors. That means that  the single-exposure precision
of  the method  is $\sim  0.04 ~  WFI@2.2m$ pixels,  or $\sim  10$ mas
(i.e.\ for each coordinate, 0.028 pixels, or 6.7 mas).

%______________________________________________________________
%
%_________________________________________________________________
%

%
% FIGX --- 
%--------------------------
\begin{figure*}[ht!]
\begin{center}
\begin{tabular}{cc}
\includegraphics[width=7cm]{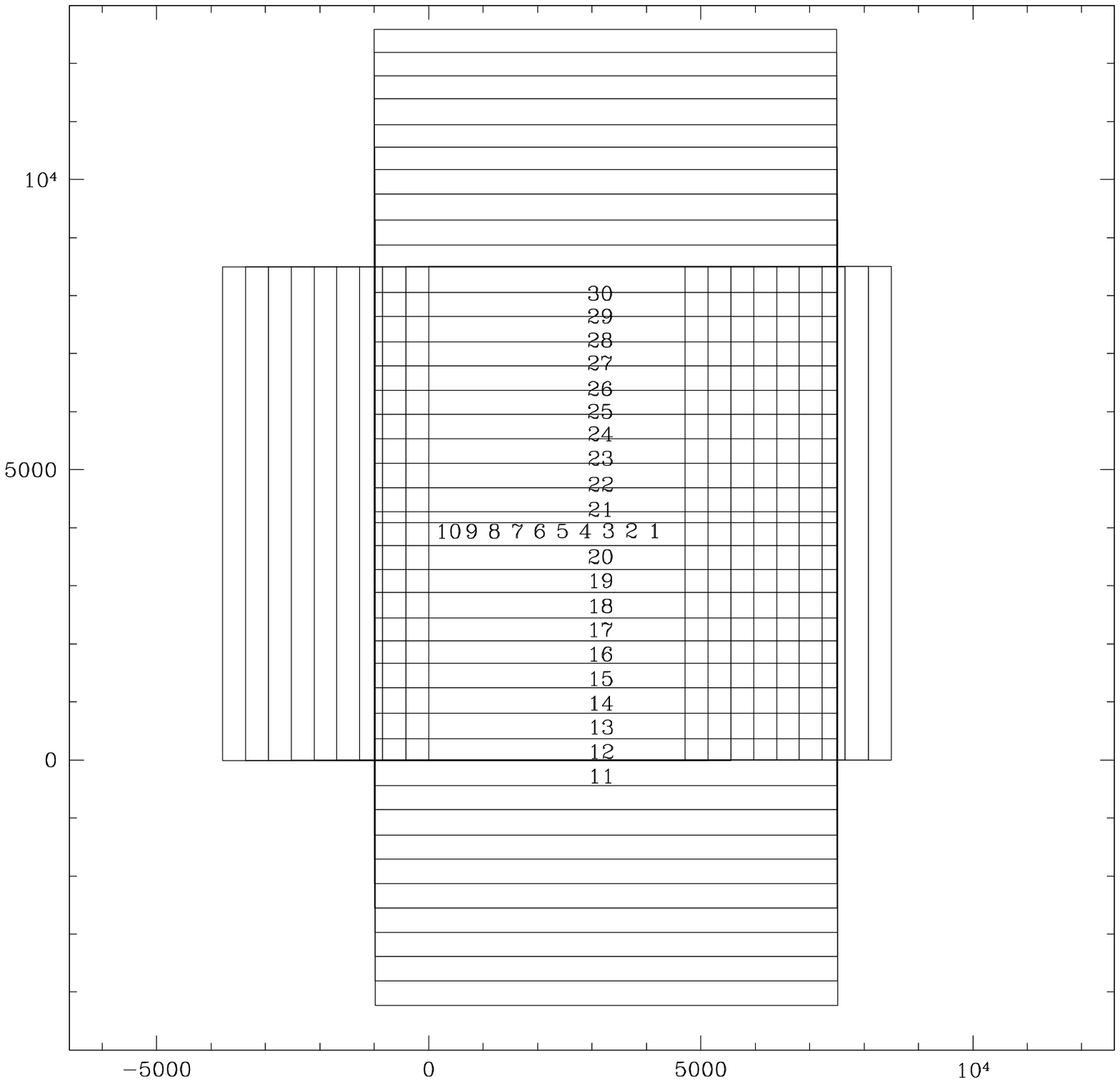}&\includegraphics[width=7cm]{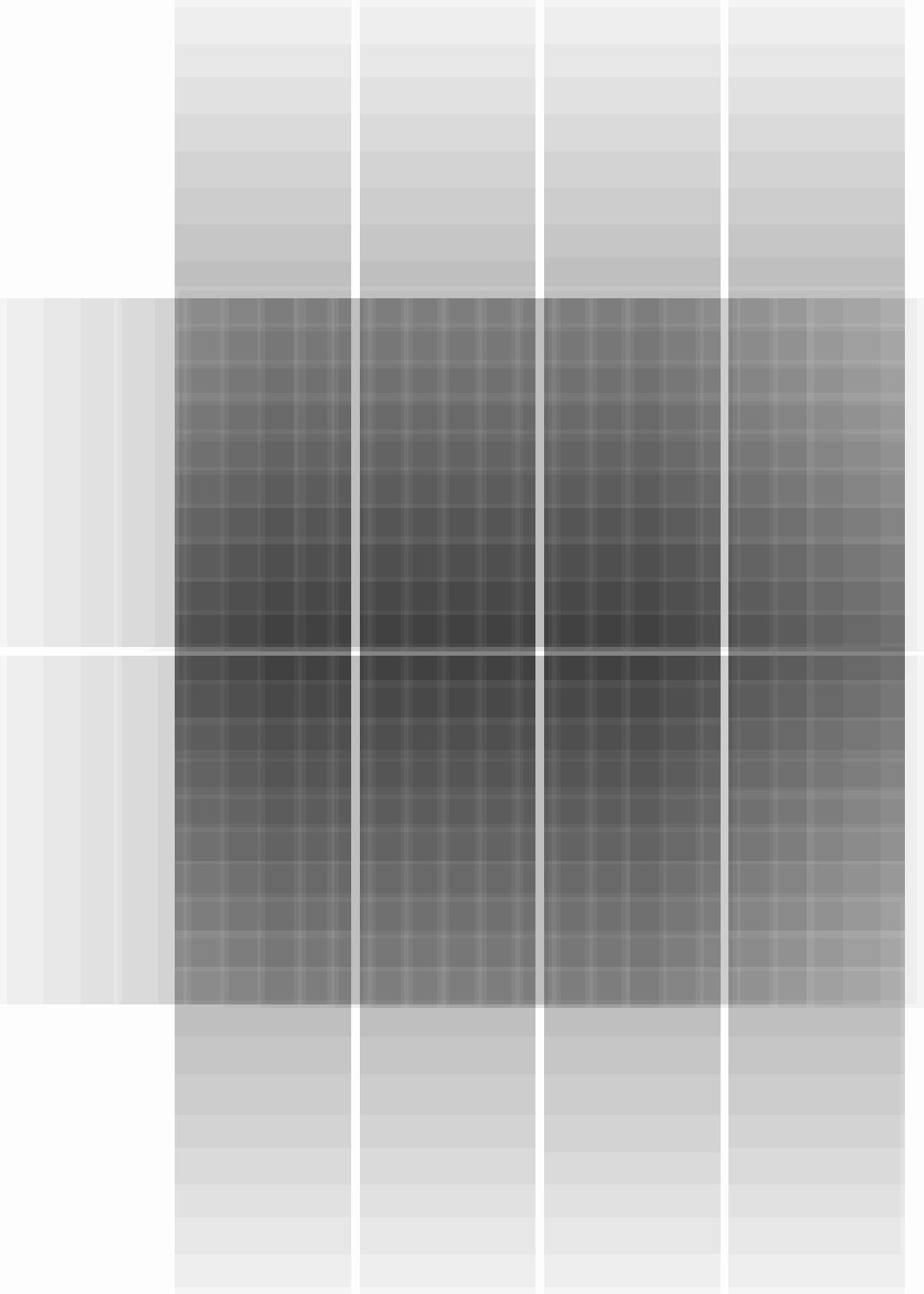} \\
\end{tabular}
\protect\caption[]{
{\em (Left)} Dither pattern of the 30 images taken in the Baade's window, in filter $V$.
	       The coordinates, are in unit of pixels, and referred to the master frame.
{\em (Right)} Depth-of-coverage map for the same 30 images.
}
\label{dither}
\end{center}
\end{figure*}
%------------------------

%%%%%%%%%%%%%%%%%%%%%%%%%%%%%%%%%%%%%%%%%%%%%%%%%%%%%%%%%%%%%%%%%
%
\section{Geometric Distortion Correction}
\label{GC}
%
%%%%%%%%%%%%%%%%%%%%%%%%%%%%%%%%%%%%%%%%%%%%%%%%%%%%%%%%%%%%%%%%%

%
As we  mentioned in  the introduction, there  is a lot  of astrometric
potential for these new wide-field ground-based images.  Although they
can  not even  approach $HST$  precision, they  can cover  an enormous
field of view, and can go  much deeper with a better accuracy than the
previous technology of photographic plates.
We saw in the previous section that we can measure a reasonably bright
star with  a precision  of about $\sim$0.03  pixel ($\sim$7 mas)  in a
single exposure.
These tests involved a local, differential measurement.  We still need
to determine what limitations distortion  will place on our ability to
measure positions in  a more global sense.  In  particular, we need to
ask: (1) how well we can  measure the distortion solution? and (2) how
stable the solution is.  The answers to these questions will determine
what kinds of astrometric projects we can undertake.

When  trying to  measure the  distortion in  an instrument,  one would
ideally like to have reference to  a list of the positions of stars in
the field in some absolute and accurate system, so that after allowing
for a linear transformation, we can see the distortion in our detector
by virtue of the position  residuals.  Unfortunately, we are not aware
of the existence of any  astrometric standard field that would allow a
direct calibration of the distortion in $WFI@2.2m$.  One might be able
to use Hipparcos
% R: hipparcos gives 120,000 stars.  
%    i.e. 3 stars for square degree. (nothing!)
%    Tycho2 2.5 million, i.e. 60/square degree.
or the USNO B survey to  calibrate the most global terms, but to fully
calibrate $WFI@2.2m$,  we would really need  a field that  is at least
0.5 degree on  a side, with tens of thousands of  stars, each of which
should have a position good to better than 5 mas.

We note that Platais et  al.  2006 proposed constructing such a frame,
and  are now in  the process  of taking  observations for  it (private
communication). Until  such a calibrated  field is available,  we must
undertake a self-calibration,  similar to what we have  done for $HST$
(Anderson \& King, 2003b).  This involves imaging a nicely dense field
with a range of telescope offsets.   Since we know that the stars have
not moved much  in the course of the night, we  can use their apparent
positions in  each of our  images to solve  for the distortion  in the
detector.

So, to do  this, as a back-up program  during a non-photometric night,
we took several  images of the Galactic Bulge,  in Baade's Window.  We
chose this field because it  is fairly homogeneous in both spatial and
star  luminosity distributions,  and  the crowding  is not  excessive.
This  is  true  across  the  entire  wide field  of  view  (see  Fig.\
\ref{baadeW}), with the exception of  the cores of two angularly small
globular clusters.  The exposure time of these images was optimized in
order to obtain the necessary number of stars with the needed S/N.

We took 30 images in the  $V$ filter, with large dithers following the
pattern shown  in Fig.\  \ref{dither}.  The idea  was to map  the same
patch of sky  into as many different places on  the 8-chip detector as
possible.
Unfortunately, the  focus was unstable, and  we had to  adjust it from
observation to observation.  Since we needed to take a large number of
images, we  did not want  to stop to  concentrate on the focus,  so we
simply made small adjustments  to the focus between observations based
on the FWHM  of the proceeding images.  This  focus variation may well
have an effect on the distortion solution, but that is part of what we
are trying to study here.

%%%%%%%%%%%
\subsection{Finding the average distortion solution}
%%%%%%%%%%%
Our first  step in solving for  the distortion was to  bring the eight
chips of  each observation into a common  meta-chip coordinate system,
using only integer-pixel shifts, in a way that the coordinates of the
pixels have  been made to correspond  as closely as  possible to their
relative locations  on the  sky.  In this  crude meta-chip  system, we
compared positions  of the same  stars in different images,  and found
that the  position residuals were  as large as 5-10  pixels---there is
clearly a lot of distortion.

The  procedure we  followed  to derive  the  correction for  geometric
distortion is an iterative  one.  We first parametrized the distortion
solution by a look-up table  of corrections for each chip that covered
each  2048$\times$4096-pixel chip,  sampling every  256  pixels.  This
resulted in a 9$\times$17 element  array of corrections for each chip.
The  distortion-corrected  position  for  a  star  will  then  be  the
meta-image   position   plus   the   interpolated   value   from   the
distortion-correction table:
%
%%%
\begin{equation}
\begin{array}{l}
X_{\rm corr} = x_{\rm meta} + \Delta x_{\rm GC}(x,y)\\
Y_{\rm corr} = y_{\rm meta} + \Delta y_{\rm GC}(x,y).\\
\end{array}
\label{eq:mu4AB}
\end{equation}
%%%
%
where $\Delta x_{\rm GC}(x,y)$  and $\Delta y_{\rm GC}(x,y)$ come from
interpolating the table for that chip.

Before  solving for the  distortion, we  first measured  positions and
fluxes for all  the stars in all the  images and cross-identified them
using a master list.
The challenge  in finding  an optimal distortion  solution is  then to
find the set  of table values which will allow  the various star lists
to be  transformed into each other using  only linear transformations.
So,  we are  looking for  a single  set of  table values  that  can be
applied to  all the images  to minimize the  non-linear transformation
residuals.

Since our aim is  differential astrometry and not absolute astrometry,
we use  general 6-parameter  linear transformations.  These  allow for
offset, rotation, and scale changes, but also allow for the axes to be
non-perpendicular  and  differently   scaled.   These  general  linear
transformations      implicitly       remove      the      first-order
atmospheric-refraction terms.

The  transformation residuals  are constructed  as follows.   For each
pair of images $i$  and $j$, we find all the stars  that are common to
both star lists.  These $N$ common stars give us $N$ pairs of position
associations: $(x_i,y_i;x_j,y_j)$.   We first correct  these positions
using   the   current   best   distortion   solution,   so   we   have
$(X_i,Y_i;X_j,Y_j)$  for each  star.   We then  find  the best  linear
transformation between the frame by least squares.

This allows  us to compute  residuals.  For each  of the $N$  stars in
image $i$,  we have  $(x_i,y_i,\delta x_i,\delta y_i)$,  where $\delta
x_j$ and $\delta  y_j$ correspond to the difference  between where the
star was found  in image $i$ and where its position  in image $j$ says
it should  be in image $i$  (based on the  linear transformation).  We
also  have   a  similar  residual  for   image  $j$,  $(x_j,y_j,\delta
x_j,\delta y_j)$.  We generate such  residuals for each star common to
each image pair.  This results in many tens of millions of residuals.

Each residual  has several contributions: the distortion  error in one
image,  the  distortion  error   in  the  comparison  image,  and  the
inevitable measurement  error.  If we examine {\it  all} the residuals
from all image pairs for a particular region of the detector, then the
average residual  will be indicative  of the distortion error  at that
chip location.   The other contributions to the  residuals will cancel
out.    Thus,   we  examine   the   residuals   about   each  of   the
distortion-array grid  points to determine how the  correction at that
grid point should  change to better approximate the  distortion in the
image.

We began the solution with  a null correction table.  We next examined
all the residuals in the  vicinity of each grid-point and adjusted the
correction at that point by  half the recommended adjustment.  We then
smoothed the  distortion table  with a 5$\times$5  quadratic smoothing
kernel to ensure that our  correction table would be smoothly varying.
At the end of the iterations  we verified that this smoothing did not
compromise our solution.

Once we have an  initial estimate for the distortion-correction array,
we solve  once again for the  residuals, but this time  we include the
correction  in the computation  of the  residuals.  The  residuals get
smaller, and  now reflect the  errors in the distortion  solution.  We
repeat  this  several times  until  we  converge  on a  final  average
solution  for  the  image   set.   Convergence  is  reached  when  the
iteration-to-iteration adjustment  for the distortion-correction array
is less than 0.005 pixel.

\begin{figure}
   \centering
   \includegraphics[width=9.cm]{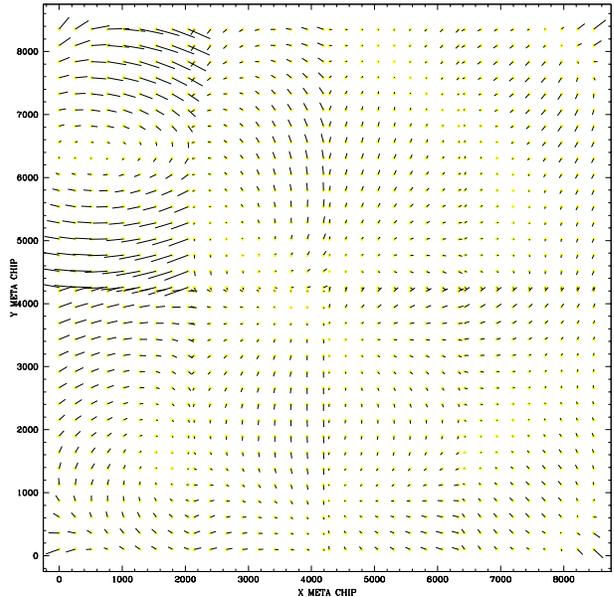}
      \caption{Map of the geometrical distortion presented in this
		paper. The corrections are exaggerated by a factor
                of 100. }
         \label{geoCOR}
   \end{figure}

The  table of  distortion corrections  is shown  graphically  in Fig.\
\ref{geoCOR}, where the corrections  have been exaggerated by a factor
of 100.  Note  that the upper left chip is  significantly rotated with 
respect to the others.

% FIGX --- 
%______________________________________________________________
%
%_________________________________________________________________
%

%%%
%
\subsection{Stability of the geometrical correction}
\label{GCstability}
%
%%%
Now that we have in  hand an average, global distortion correction, we
can investigate how it may change over time, both over the course of a
night  and  in  the longer-term.   To  do  this  we  took the  30  $V$
observations  above  and  generated   a  master  frame  based  on  the
centermost  dither pointing.   We then  transformed each  image's star
positions into this frame, and  arrived at an average position for the
stars in this frame.  This average  frame allows us to look at how the
solution may vary over the course of a night.  Even though our average
frame  converged  to  better  than  0.005  pixel  we  found  that  the
individual frames can have  residual distortions as large as $\sim$0.2
pixel (i.e.\  $\sim$50 mas).   Much of  this is near  the edge  of the
field  and is  likely due  to telescope  flexure or  focus variations.
This  would be  extremely hard  to model  predictively, and  it likely
would change as conditions change.

Comparing two  different epochs, separated  by $\sim$3 years,  we find
that the geometric  distortion can vary up to  0.4 pixel, or $\sim$100
mas.   Of  course, since  the  2.2m  ESO/MPI  telescope is  not  fully
dedicated to $WFI@2.2m$,  and other instruments can be  mounted on the
same  telescope, manipulation  of the  camera can  easily result  in a
large variation  of the geometrical distortion. In  particular we know
from other data sets that a rotation  of the whole camera by up to few
degrees may be present from run to run.

We  also  found in  our  data  set that  filters:  $V$,  $B$, $I$  and
$I_{\#853}$  share the same  distortion solutions  down to  $\sim 0.4$
pixels. Filter $U$  is an exception, here the  differences may rise up
to 5 pixels  close to the edges  of the field of view.  In the present
work we are not using the $U$ filter for astrometry. We will calibrate
the geometric solution  in this filter when a  better data-set becomes
available.

Thankfully, our  proper-motion membership measurements  do not require
us to  know the  distortion solution perfectly.   We can  minimize the
impact of  distortion errors by using local  transformations, which we
discuss in Section 7.

%__________________________________________________ 
% 

%_________________________________________________________________
%

%%%%%%%%%%%%%%%%%%%%%%%%%%%%%%%%%%%%%%%%%%%%%%%%%%%%%%%%%%%%%%%%%
% FIGX --- 
%______________________________________________________________
%
   \begin{figure}[!ht]
   \centering
   \includegraphics[width=9.cm]{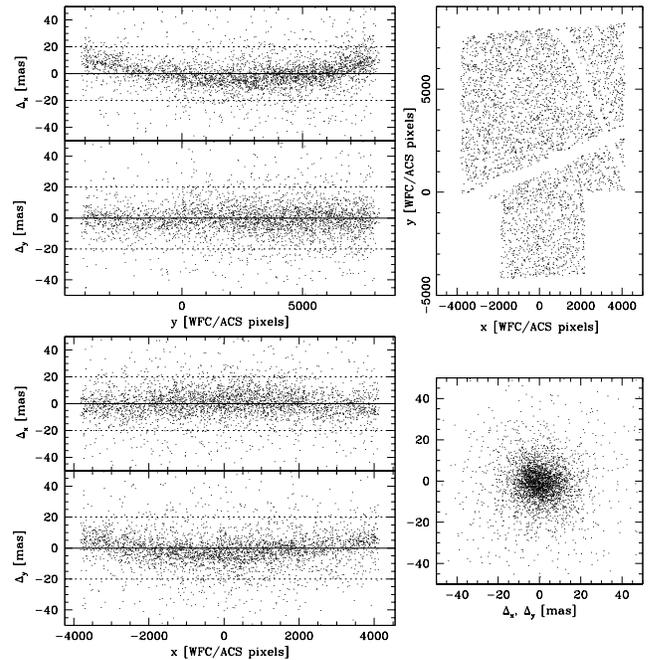}
      \caption{
%	%
%	%
	{\em (Left  Panels):} We show the residuals  in the distortion
	solution  for  $y$ and  $x$  vs.\  $dx$  and $dy$.  Note  that
	residuals are smaller than 20 mas in the common area.
%	%
%	%
	{\em (Top Right):} Spatial distribution of stars which have 
	measured positions in both the $HST$ and $WFI@2.2m$ data base
	(one WFC/$HST$ pixel is 50 mas). 
%	%
%	%
	{\em (Bottom Right):} Residuals of positions measured in the
 	$WFI@2.2m$ images and transformed into the $HST$ master frame
 	to cross-check the solutions.
%	%
      }
         \label{hst}
   \end{figure}
%_________________________________________________________________
%

%%%%%%%%%%%%%%%%%%%%%%%%%%%%%%%%%%%%%%%%%%%%%%%%%%%%%%%%%%%%%%%%%
%
\section{Comparison with $HST$}
\label{HSTcomp}

Although the ACS/WFC  field of view is only  202$''$ $\times$ 202$''$,
which is just $\sim$1\% that of the $WFI@2.2m$, we can still use $HST$
observations  as  a  cross  check  on our  measurement  precision  and
distortion correction.

As  luck would have  it, $HST$  happened to  observe the  same Baade's
window field as  we have been analyzing here within a  few days of our
observations (GO-9690).  With the typical motion of a bulge star being
$\sim$3 mas/yr  (Bedin et  al.\ 2003b, Kuijken  \& Rich 2002),  we can
assume that  the stars are  all in the  same place, and treat  the WFC
images as a  perfect reference frame against which  we can compare our
reductions.   The  footprint  of   this  5  field  WFC/ACS  mosaic  is
superimposed on one WFI image in Fig.~\ref{baadeW}.

We reduced  the WFC images as  described in Bedin et  al.\ (2005b) and
tied them  together, obtaining  a huge distortion-free  frame covering
over 50 contiguous square arcminutes  with a global precision of about
1 mas (see  Anderson \& King 2006).  This represents  about 5\% of the
$WFI@2.2m$ field of view.

So, to compare against the  $HST$ frame, we cross-identified the stars
from our average  frame in the previous section with  the stars in the
$HST$ list.  We then  transformed the $WFI@2.2m$ measurements into the
$HST$  frame  using a  global  linear  transformation  to construct  a
measurement  residual for each  star.  Since  the $HST$  positions are
much  better  measured  than   those  on  ground-based  images,  these
residuals  will   provide  a  fundamental  test   for  the  $WFI@2.2m$
astrometry.

Fig.\ \ref{hst}  shows the residuals.  In  the 4 boxes on  the left we
show the behavior of the residuals  in $x$ and $y$ as function of both
the  coordinates.   In  the  top  right  panel  we  show  the  spatial
distribution of  the stars on  the WFC/ACS $HST$-mosaic  master frame.
Note  that there  are  gaps, which  are  caused by  the  fact that  we
required stars  to be  detected in 18  out of  the 30 $V$  images; the
dither pattern  ended up  placing the  same star in  the gap  for many
exposures.   The bottom  right  panel shows  the  distribution of  the
residuals.

The fact that the solution  appears to get worse over larger distances
is  consistent with  our  finding  in the  previous  section that  the
solution is not stable at the  0.2 pixel level (50 mas), and that most
of the  variation comes  from low-order terms  which are worse  at the
edges.  This independent test shows that over distances of 5$'$ or so,
an average  frame produced  with our corrections  is accurate  to much
better  than  20 mas  ($\sim$0.08  $WFI@2.2m$  pixel).  However,  this
figure and  Fig.\ \ref{rms1}  show that we  can clearly  measure stars
with random errors of less than  10 mas (i.e.\ 0.04 pixel), so we will
have  to  find  some  way  to minimize  these  global  errors.   Local
transformations will provide that means.

%%%%%%%%%%%%%%%%%%%%%%%%%%%%%%%%%%%%%%%%%%%%%%%%%%%%%%%%%%%%%%%%%
%
\section{Local-transformation approach}
\label{LOCAL}
%
%%%%%%%%%%%%%%%%%%%%%%%%%%%%%%%%%%%%%%%%%%%%%%%%%%%%%%%%%%%%%%%%%

%%%%%%%%%%%
\subsection{The reference-frame problem}
%%%%%%%%%%%

We saw  in the previous section  that the distortion is  not stable at
the level of our  intrinsic astrometric accuracy ($\sim$0.03 pixel, or
$\sim$7 mas,  in each coordinate).  This  means that if we  want to do
high-precision astrometry  with our images, we  need to find  a way to
minimize  the  effect  of  uncorrected  distortion.  This  led  us  to
investigate a local approach: differential astrometry.

If we want to measure how a  star has moved from one epoch to another,
we always need to have a frame of reference, in which we can compute a
position  at one  epoch  and a  position  at the  other  epoch, and  a
resulting displacement.  Now,  we do not require an  absolute frame of
reference  here.  There are  very few  stars in  a typical  field with
absolute positions and motions that are known to a useful accuracy.

Thus, our reference  frames will necessarily be relative.   By this we
mean that we will measure our motions with respect to something in our
field.  We could just use all the stars as a reference and our motions
would  be  with  respect to  the  average  motion  of all  the  stars.
Typically, however, we  will pick a particular population  and set its
motion to zero and measure motions relative to that population.  Since
cluster stars  tend to  have less internal  dispersion than  the field
stars, it  is natural to  use the cluster  as the reference.   We will
thus identify the obvious  cluster-member stars (either by location in
the CMD, or  iteratively by the motion itself, or  both), and use them
as the basis for the transformation.

Each star's  motion, then, will be  measured with respect  to the bulk
motion of the cluster.  Cluster-member stars will by construction have
no motion  on average, but if  our measurements are  precise enough to
resolve  the  cluster dispersion,  this  dispersion  will  show up  as
residuals from the average.  In  the outskirts of clusters, the member
density is  often extremely low.  In  these regions, there  may not be
enough members in each chip to serve as reference stars.  Thus, it may
be necessary to base the  transformations on the non-member stars.  In
this case,  the cluster may show  up as a concentration  in the motion
diagram  that is  not necessarily  at the  origin.  But  this  was not
necessary  for the  the present  work; even  in the  outskirts  of the
clusters there were plenty of member stars to use as reference stars.

We saw in  the previous section that all the frames  of our images may
contain some uncorrected  distortion, so we will not  have access to a
flat global  reference frame at  any epoch.  Rather  we have a  set of
frames,  one  for each  image  in each  epoch.   Each  frame has  been
corrected for distortion  as well as possible, but  it will invariably
have some  residual distortion.  This residual distortion  tends to be
of  a  global  nature,  in  that  we can  compare  nearby  stars  more
accurately than we can compare stars that are farther apart.  Thus, if
we confine our  comparisons to a small region about  each star, we can
minimize the effects of distortion errors.

In images with  $HST$, the distortion errors are  small enough that we
can generally  define a  decent global reference  frame and  use local
transformations to improve positions  in this frame.  With $WFI@2.2m$,
the  distortion errors  are  larger,  and it  is  more problematic  to
generate an  accurate master  frame.  For this  reason, we  decided to
treat  each frame independently  and compute  a displacement  for each
star for  each pair  of frames.  If  we have  (say) 10 frames  in the
first  epoch and 9 frames in  second epoch,  this gives  us 90
estimates of  the inter-epoch  displacement.  These estimates  are not
all statistically independent of each other, but simple statistics can
tell us how to combine them and estimate the error in the average.

So, there  are two steps in  our construction of  proper motions.  The
first step is  measuring a displacement between each  image pair.  The
second step is  combining the many measurements in  order to obtain an
average displacement from one epoch to another.

%%%%%%%%%%%
\subsection{Step 1: measuring displacements between a pair of images}
%%%%%%%%%%%
%
The  first  step  in  measuring  a  proper  motion  is  to  compute  a
displacement for  a star that is  measured in two images  taken at two
different epochs.   We start here  with a list  of $N$ stars  that are
found  in both  images.  We  have  a position  for each  star in  both
distortion-corrected frames: $(x1_n,y1_n)$  and $(x2_n,y2_n)$.  We can
use  these  associated positions  to  define  a linear  transformation
between   the  frames,   and  compute   a  global-transformation-based
displacement  for   each  star.   This   displacement  represents  the
difference between  where the star was  measured to be  in Frame~1 and
where  the Frame~2 position  implies it  is in  Frame~1 (based  on the
positions of the common stars).   If there is no distortion, then this
displacement  will be  a  good estimate  of  the actual  displacement.
However,  if there is  distortion in  either of  the frames,  then the
estimate could contain some serious systematic errors.

Since we  know that  the $WFI@2.2m$ frames  suffer a  distortion error
that increases  over larger  distances, we can  minimize the  effect of
uncorrected  distortion  on  our   displacements  if  we  measure  the
displacements    using   local   transformations.     Whereas   global
transformations  use  all  the   stars  to  compute  a  single  linear
transformation  between two  frames, a  separate  local transformation
will  have to  be computed  for each  star we  wish to  transform from
Frame~2 to Frame~1.

We will compute this transformation using a local subset of stars.  To
do  this,  we find  the  closest 55  reference  stars  to the  Frame~1
position, choosing  only stars in  the same chip  as the star  in both
images.   The positions  of  these stars  in  both frames  give us  55
associations,  positions in  one  image that  correspond  to the  same
position in another image.  We  can use these associations to define a
least-squares linear  transformation from one frame to  the other.  We
are careful not to use a star in its own transformation, as that would
introduce a bias  in its displacement (the bias  would reduce the true
displacement by 1/55).

\begin{figure}[!ht]
   \centering
   \includegraphics[width=9.cm]{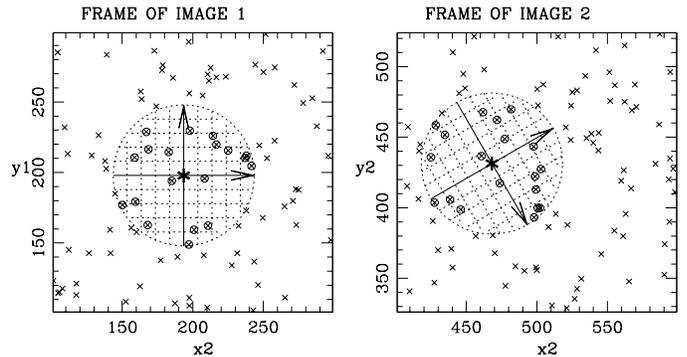}
      \caption{Visualization  of  the  local-transformation  approach.
        Consider   Frame1  and  Frame2,   observing  the   same  stars
        (indicated  by $\times$s) but  with different  orientation and
        shift.  We  are trying to  transform the position of  the star
        \textbf{\Large $\ast$} measured in  Frame2 into Frame1. In the
        local  approach  only  the  closest  stars  (highlighted  with
        $\circ$)   are  used   to  define   the  6   parameter  linear
        transformations.}
         \label{net}
   \end{figure}

Figure~\ref{net}  shows how the  positions of  reference stars  in two
frames  can  allow  us  to  define  a  transformation  that  maps  the
coordinate system  of one frame into  that of another, so  that we can
transform a  position measured  in one frame  into that of  the other.
For clarity,  in this figure we  only show a few  reference stars.  In
our transformations, we use 55 neighbors.

Now, not  all of these stars are  as good as the  others at specifying
this  transformation.  For  one, even  though we  chose  our reference
stars to be  members based on their location in the  CMD, some of them
may not be moving with the reference frame, so we automatically reject
all  stars that  are  not moving  with  the ``reference  population''.
There are  also some stars that may  be poorly measured in  one of the
frames, or may  have an anomalous motion.  We do  not want these stars
to bias our transformation, so we iteratively reject the 10 stars that
have the largest transformation residuals.

Once  we   have  the  45  best   local  stars,  they   will  define  a
transformation that will allow  us to transform $(x2_n,y2_n)$ into the
first  frame.  We  can then  compute a  displacement  trivially.  This
displacement, divided by the  time baseline, provides one estimate for
the  proper motion.   This estimate  is based  solely on  positions of
stars in the two images being compared.

%
% FIGX --- 
%______________________________________________________________
%
%_________________________________________________________________
%

%%%%%%%%%%%
\subsection{Step 2: combining the measurements from all pairs}
%%%%%%%%%%%

Using the technique in the  previous section, we can compute a locally
measured displacement  for each star for each  inter-epoch image pair:
$(\Delta x_{ij,n},\Delta y_{ij,n})$, where  $i$ is a first-epoch image
and  $j$  is  a   second-epoch  image.   If  there  are  $\mathcal{I}$
first-epoch images and $\mathcal{J}$ second-epoch images, then we have
$\mathcal{I} \times \mathcal{J}$ measured displacements for each star.
These  displacements  are not  statistically  independent, since  each
first-epoch image $i$ uses  the same $\mathcal{J}$ second-epoch images
to construct its displacements.

If we can assume that all  the observations are good, then we can take
a simple average of  these displacement measurements.  This average is
the best  estimate for the inter-epoch  displacement.  Determining the
error in this is trickier.

To estimate  the error  in our displacemets,  we need to  estimate the
error in each first and second epoch position.  We compute intra-epoch
displacements  in a  similar  manner to  the  above {\it  inter}-epoch
displacements by taking all pairs  of images from the same epoch.  The
r.m.s.\  of   these  displacements  (the  average  will   be  zero  by
definition)  will  give  us  an  estimate  of  the  accuracy  of  each
individual   measurement   for    each   epoch,   $\sigma_{x_1}$   and
$\sigma_{x_2}$.  The error  in the average epoch 1  position from this
would then be:
$\sigma_{\bar{x1}} = \sigma_{x_1} / \sqrt{\mathcal{I}} $ and the error
in the epoch 2
position  $\sigma_{\bar{x2}} =  \sigma_{x_2}/\sqrt{\mathcal{J}}$.  The
error in the difference would then be:
$$
      \sigma_{\Delta x} = \sqrt{\sigma_{\bar{x_1}}^2/\mathcal{I}  + 
                                \sigma_{\bar{x_2}}^2/\mathcal{J}  }.
$$
This gives us the error in the average.

We can  also use the intra-epoch  analysis to determine if  any of the
measurements from an  epoch are inconsistent with the  others.  We can
look at  all the  residuals for image  $i$, and  if there is  one much
larger than  $\sigma_{x1}$, then we  can reject this  observation, and
recompute  this with  $\mathcal{I}-1$ first-epoch  measurements.  This
will lead to a more robust average displacement.

We note that in many projects, we  will not need to know the errors of
our motion measurements precisely.  It is often clear from the diagram
itself that we have effected a near-perfect separation between cluster
and field, and  a precise understanding of the  errors is unnecessary.
When we are  trying to measure internal motions  (dispersions), then a
proper understanding of the errors will be crucial.

Also, there are some data sets  that will have more than two epochs of
observation.   We   can  use  the   above  techniques  to   compute  a
displacement for each star for each pair of epochs.  We can then use a
similar approach  to combine these displacements into  a single proper
motion.

%-----------------------------------------------------------------
% 
% 
%
% FIGX --- 
%______________________________________________________________
%
   \begin{figure*}
   \centering
   \includegraphics[width=18.cm]{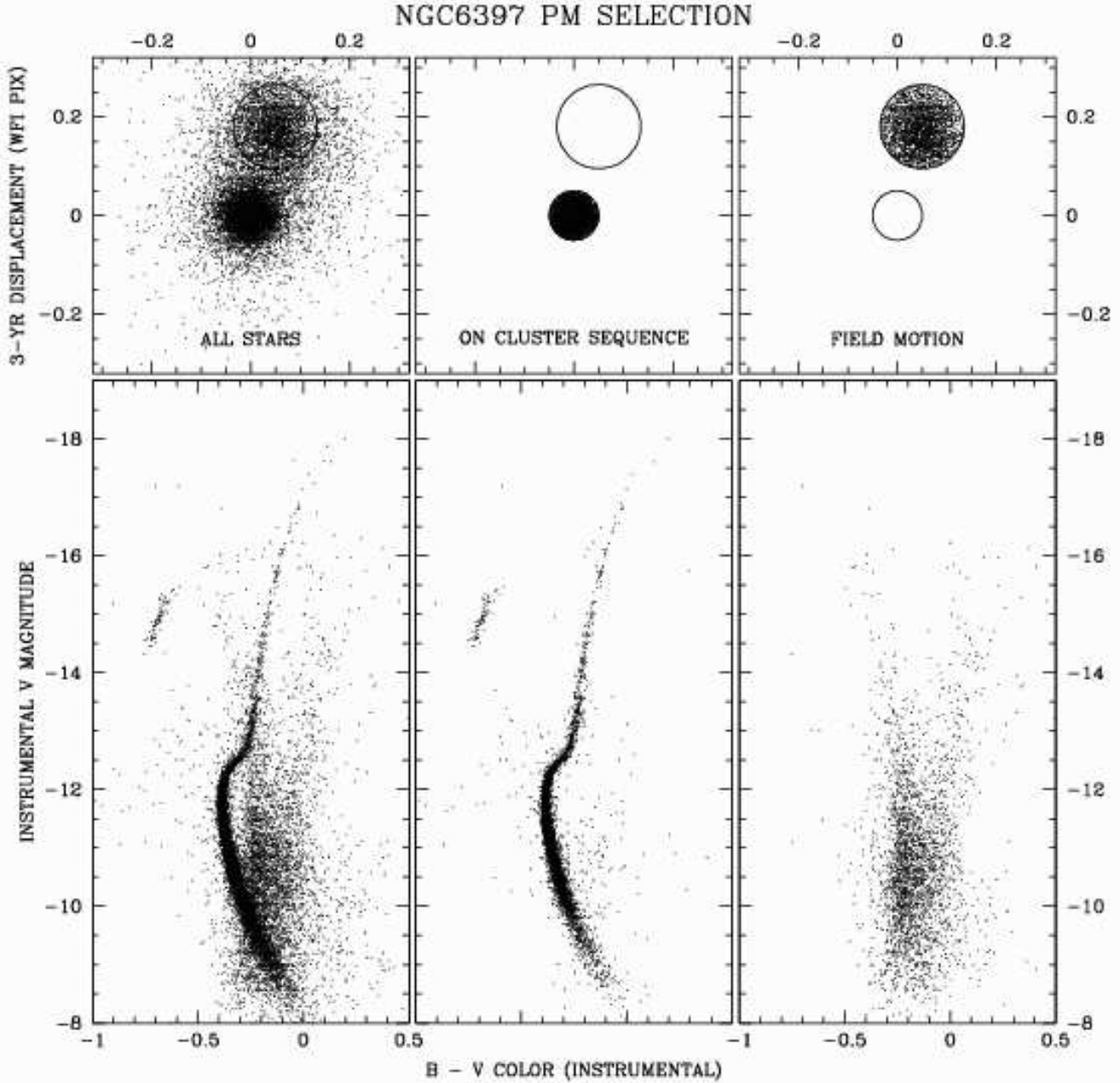}
      \caption{
%	%
	{\em (Top  panels)} Vector point diagrams  of displacements of
	stars in units of  $WFI@2.2m$ pixels (238 mas/pixel) after 3.1
	years.  Since  all the  reference stars were  cluster members,
	the zero point of the motion is the mean motion of the cluster
	stars.
%	%
	{\em    (Bottom   panels)}   Instrumental    color   magnitude
	diagrams. The magnitude is  calculated as $-2.5 \log{\rm DN}$,
	where DN is  the total digital counts above  the local sky for
	the considered stars.
%	%
	{\em (Left)} The entire  sample; {\em (Center)} stars with the
	same  proper motion  (within  0.05 pixels)  as  the MS  stars,
	i.e.  with  proper  motions  smaller than  3.8  mas/yr.   {\em
	(right)}  The  stars  that  fell  in the  bulk  of  the  field
	distribution.
%	%
	All these plot shows only stars with rms in positions inferior
	to 0.075 pixels in each coordinate.  }
%       %
         \label{n6397}
   \end{figure*}
%_________________________________________________________________
%

%
% FIGX --- 
%_________________________________________________________________
%
   \begin{figure*}
   \centering
   \includegraphics[width=18.cm]{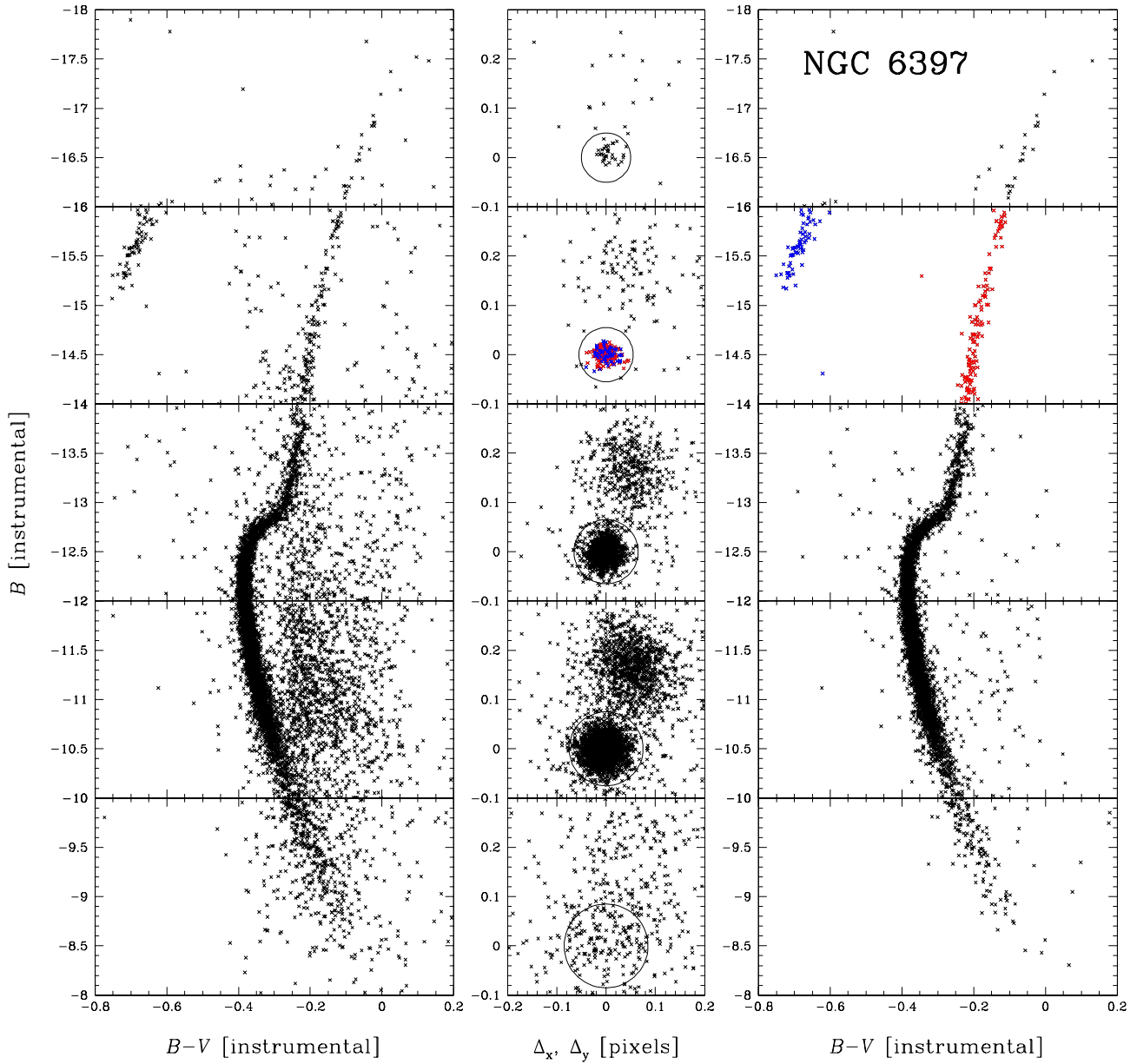}
      \caption{
	{\em (Left:)}   Color-magnitude diagram for all the stars with 
	positional r.m.s.\ less than 0.05 pixels.  
%	%
 	{\em (Middle:)}  Vector point diagrams  for the same  stars in
	the corresponding  magnitude intervals. A circle  in each plot
	shows  the  adopted membership  criterion  for that  magnitude
	interval: from top to  bottom: 0.050, 0.055, 0.065, 0.075, and
	0.085 pixel in 3 years (i.e.  3.8, 4.2, 5.0, 5.8, 6.5 mas/yr).
	Note that  the proper motions  have been corrected for  DCR as
	described in Section 9.
%	% 
 	{\em (Right:)}  Color-magnitude diagram for  the stars assumed
	to be members.  
%       %
        }
         \label{n6397_II}
   \end{figure*}
%_________________________________________________________________
%

%
% FIGX --- 
%_________________________________________________________________
%
   \begin{figure*}
   \centering
   \includegraphics[width=18.cm]{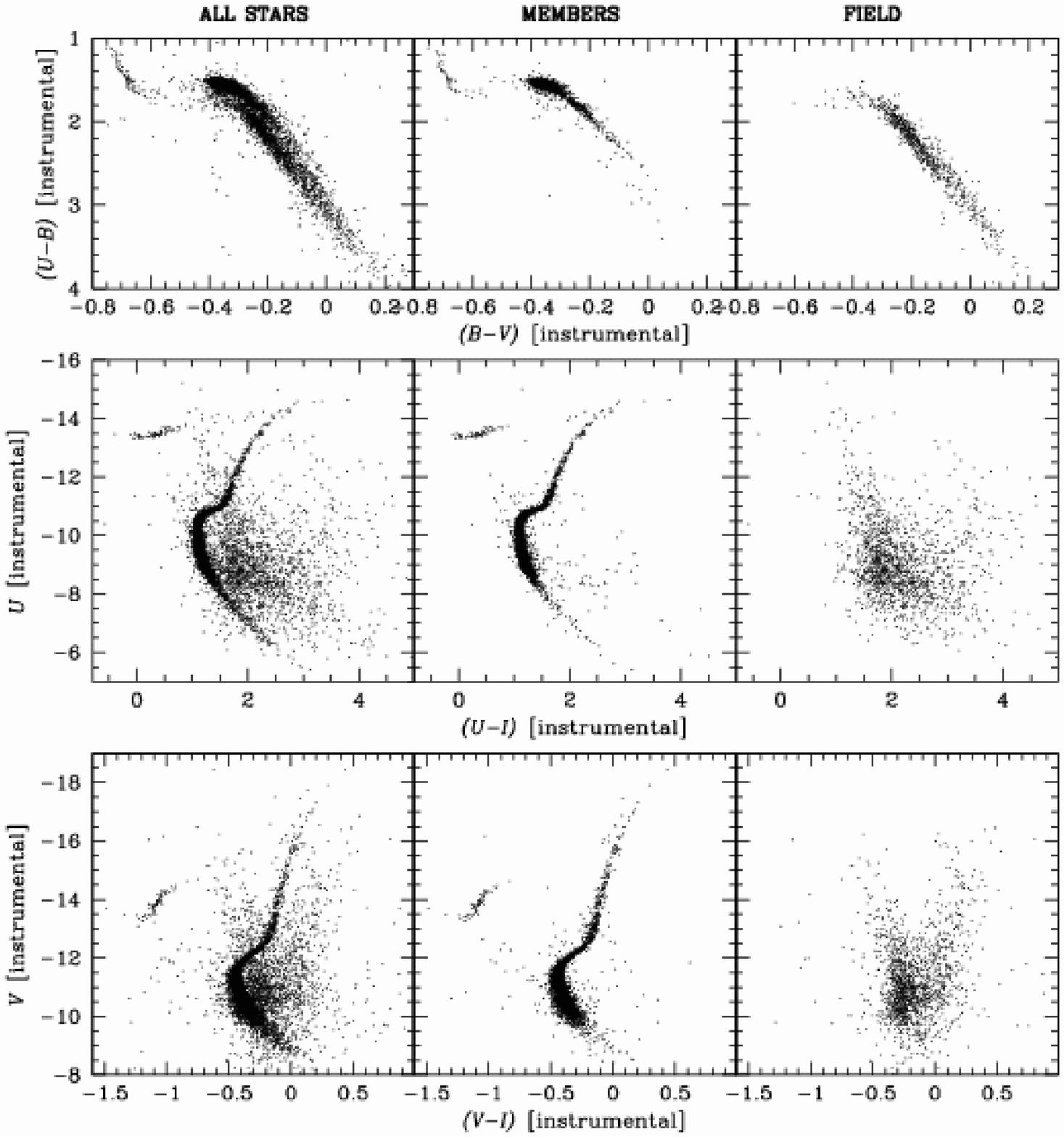}
      \caption{
%	%
        Additional  color-magnitude   and  color-color  diagrams  from
        NGC6397  WFI reductions and  the proper-motion  selection from
        the previous figures.
%	% 
	%
	}
         \label{n6397_III}
   \end{figure*}
%_________________________________________________________________
%

% 
%
%-----------------------------------------------------------------

\section{Application}
\label{APPLICATION}
As a demonstration the science that can be done with these techniques,
we downloaded  multi-epoch observations of NGC~6397  and NGC~6121 (M4)
from the ESO  archive.  These two clusters have  eccentric orbits, and
at the present day they are in a phase where their spatial velocity is
significantly different  from the  average velocity of  field objects.
They are also  the two closest clusters to us,  so the relative motion
between cluster  and field is particularly high.   Their nearness also
means that  they are best  imaged with a large  field-of-view detector
($\sim$1 degree).

For these reasons, these two  clusters are ideal targets to illustrate
what can be  done with wide field ground-based  astrometry.  We should
note that  these images  were not taken  with astrometry in  mind.  An
ideal  astrometric data set  would have  multiple observations  of the
same field  with a large number of  dithers at each epoch,  so that we
can randomize any distortion errors and improve our random measurement
errors by  $\sqrt{N_{\rm obs}}$.   These observations were  taken with
very little  dithering, so in some  sense, they represent  what can be
done with the  typical archival data set.  A  properly dithered set at
one  (or   preferably  both)  epochs  would  allow   to  optimize  the
astrometric measurements.

%%%%
%%
%
\subsection{NGC~6397}
\label{NGC6397}
%
%%
%%%%

Table~\ref{obs} lists the data available for NGC~6397.  We reduced the
images for all four filters ($U$, $B$, $V$, and $I$), but only use the
$B$ and $V$ images for proper-motion analysis.  The time baseline is a
little over three years.

The first  thing to  do is determine  which stars  we will use  in the
transformations.  Since  we have multi-color observations,  we can use
the CMD to select stars  that are likely cluster members.  We selected
stars along the main sequence  and red giant branch (RGB), leaving out
a  few cluster  stars (such  as blue  stragglers  or horizontal-branch
stars), a  fact that will not  hamper our results.   Also, including a
few field  stars that  happen to  lie on the  cluster sequence  do not
affect our measurements, thanks to the rejection criteria in our local
transformations.

We started with a list of probable members and used their positions in
the images of both epochs to transform (via local transformations) the
position of  every star in each  second-epoch image into  the frame of
the  first-epoch  image.   We  then averaged  these  displacements  as
discussed above  to arrive at a  displacement for each  star.  In this
way, we obtained a set of  diagrams similar to those plotted in Figure
11, which shows a vector-point  proper motion diagram, and the members
and non-members  selection criteria with the resulting  CMDs.  We then
iterated one more time, using as cluster members only those stars that
(a)  satisfy the  above  CMD  criterion and  (b)  have a  cluster-like
motion.  This  gives us  our final motions,  which are shown  in Fig.\
\ref{n6397}.

Fig.\ \ref{n6397} shows the results  in an analogous form to the Fig.\
1 presented by King et al.\ (1998) for space-based observations of the
same cluster (for comparison, in  that work the time base-line was 2.7
years). The cluster-field separation here is nearly as good as the one
achieved with  $HST$ with the same  baseline.  Of course we  do not go
nearly as faint here, but we cover a larger area.

In Figure  \ref{n6397} we drew  the circles which isolate  the cluster
and field stars by eye.  We  defined as a cluster member a stars which
lies  within 0.05 pixels  (i.e.\ 3.8  mas/yr) from  the origin  of the
proper motion axes.  The radius has been chosen as the best compromise
between  losing members  with  poor proper  motions measurements,  and
including  field objects.   The  internal velocity  dispersion of  the
stars in NGC 6397 is expected to be about 5 km/s, or 0.5 mas/yr at the
distance of $\sim$2 kpc.  This  should actually be measurable from the
ground with  a 10-year baseline,  or alternately more  observations at
each epoch.

Since the  fainter stars  are generally less  well measured,  we often
draw   a  more  generous   circle  for   them.   Figure~\ref{n6397_II}
illustrates  this.   (The proper  motions  in  this  figure have  been
corrected for  differential chromatic refraction  effects as described
in Section\ \ref{ATM}).

The success  of the separation  is immediately evident.   The relative
average motion  of the field with  respect to the  members is $\sim$14
mas/yr.  This  is not the absolute  proper motion of  the cluster, but
only the motion  relative to the bulk motion of the  field.  To get an
absolute  motion we should  either measure  the cluster  proper motion
with respect to  background galaxies, or the field  proper motion with
respect to background galaxies.

Note how the  membership is well established even  for saturated stars
whose  instrumental  magnitudes   are  brighter  than  $V\simeq-14.5$.
Instrumental  magnitude  is  defined  here  as  $-2.5\times  \log{DN}$
(Digital  Numbers).  With our  seeing, the  central pixel  contains no
more  than  8   percent  of  the  light,  and   saturation  starts  at
$\sim$55,000  DNs.  Therefore,  saturation begins  at  an instrumental
magnitude  of $\sim  -14.5$ [$=  -2.5 \log{(55,000/0.08)}$],  for both
$B$, and $V$.
Since our astrometry  is done locally, and we need  a dense network of
reasonably bright stars, it is hard  to use the short exposures in the
astrometric  analysis.   However,  we  can  use them  easily  for  the
photometry, so for the saturated stars we show the position in the CMD
from  the  short  exposures  (with  a zero-point  to  match  the  long
exposures) but adopt the astrometry  from the deep exposures.  (In the
next section,  we will see that image-motion  can significantly affect
astrometry in short exposures.)

In Fig.\  \ref{n6397_III} we  show how proper  motions can be  used to
clean-up  other classical  diagrams used  in the  analysis  of stellar
populations.  The top panels show the color-color $(U-B)$ vs.\ $(B-V)$
diagram,  the  middle panels,  the  color-magnitude  diagram with  the
largest color  base-line $U$ vs.\  $(U-I)$, and finally on  the bottom
the reddest color-magnitude diagram, the $I$ vs.\ $(V-I)$. 

%%%%
%%
\subsection{M4 (NGC~6121)}
\label{M4}
%%
%%%%

\begin{figure*}
   \centering
   \includegraphics[width=18.cm]{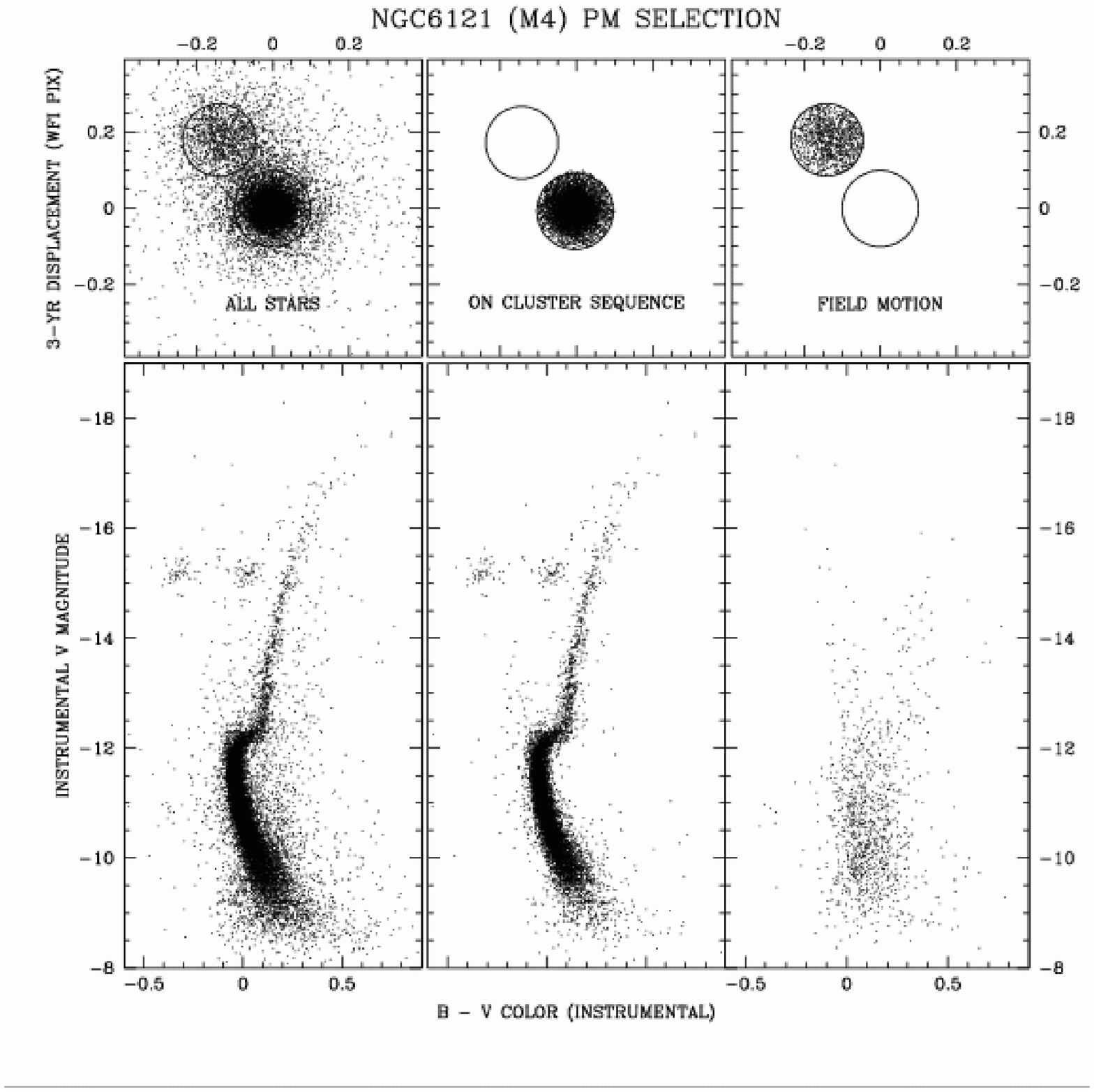}
      \caption{ 
%	% 
        As  in Fig.\  \ref{n6397} but  for  M4.  Stars  with the  same
	proper motion  of the  MS stars within  0.08 pixels  (2.8 year
	baseline), are  considered members (i.e.\  the circle centered
	in the origin includes  stars with proper motions smaller than
	6.8 mas/yr).
%	%  
	All these plots  show only stars with rms  errors in positions
	smaller than 0.05 pixels in each coordinate.
%       %
        }
        \label{n6121}
   \end{figure*}

\begin{figure*}
   \centering
   \includegraphics[width=18.cm]{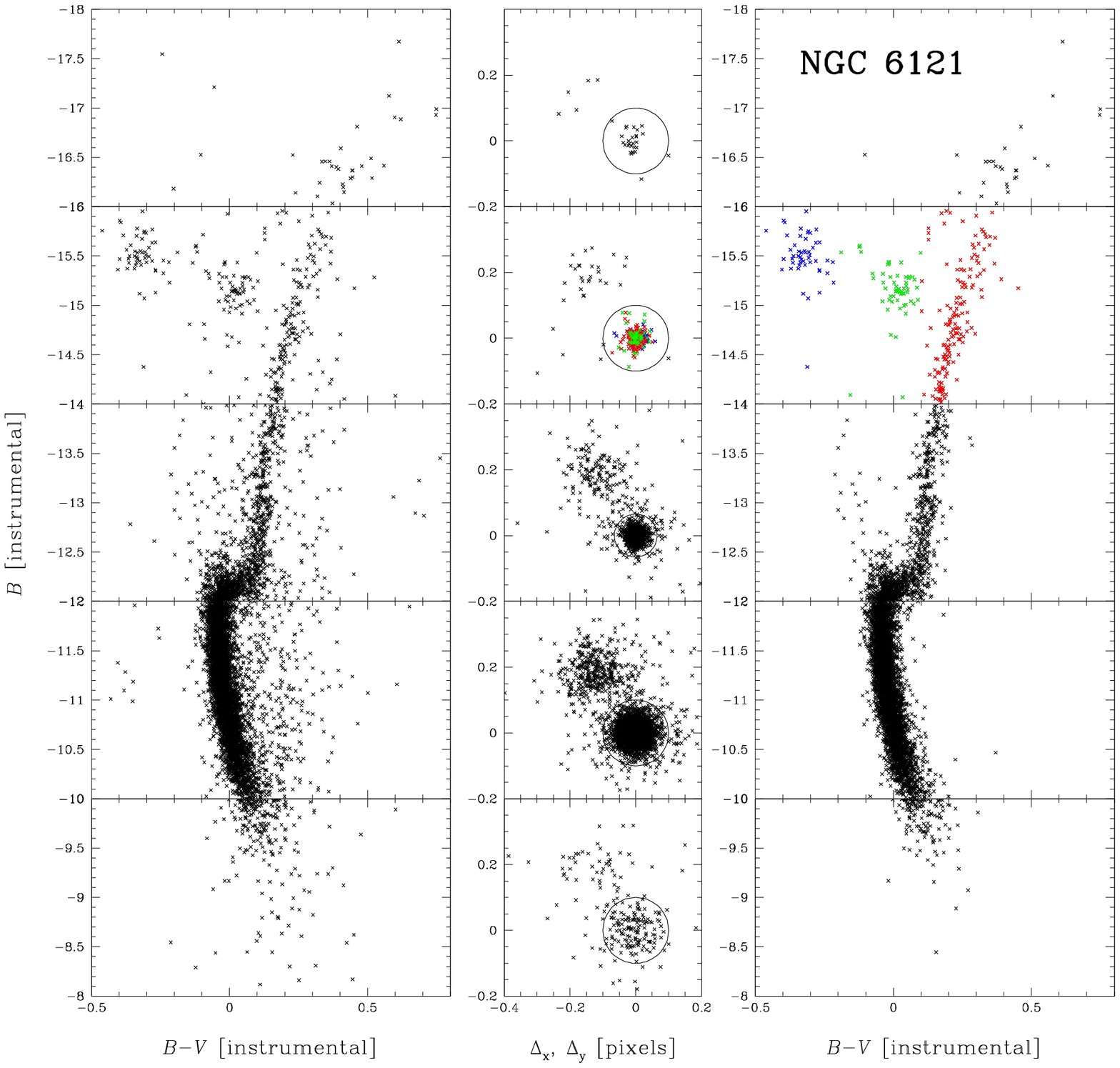}
      \caption{
	As in Fig.\ \ref{n6397_II} but for M4. 
%	%
	A circle  in each plot shows the  adopted membership criterion
	for  that magnitude interval:  from top  to bottom:  0.1, 0.1,
	0.065, 0.1, and 0.1 pixels.
%	%
	}
         \label{n6121_II}
   \end{figure*}

The same exercise has been repeated for the case of M4 (NGC~6121).  In
this case, the time base-line is just 2.8 years (see Tab.\ \ref{obs}).
The separation  based on the proper  motion is slightly  worse than in
the  case of  NGC~6397,  mainly because  of worse-than-average  seeing
(which also  affects the photometry). Differential  reddening may also
be responsible for some of the color broadening.
Nevertheless, even  with a  smaller baseline and  poor seeing,  we can
still  successfully separate  field from  cluster stars  (cfr.\ Figs.\
\ref{n6121},  and   \ref{n6121_II}).   The  adopted   radius  for  the
membership circle is  of 0.1 pixels, corresponding to  a proper motion
of 8.5 mas/yr.

% FIGX --- 
%______________________________________________________________
%
%_________________________________________________________________
%
%
% FIGX --- 
%_________________________________________________________________
%
%_________________________________________________________________
%

%%%%%%%%%%%%%%%%%%%%%%%%%%%%%%%%%%%%%%%%%%%%%%%%%%%%%%%%%%%%%%%%%
%
\section{Atmospheric Effects}
\label{ATM}
%
%%%%%%%%%%%%%%%%%%%%%%%%%%%%%%%%%%%%%%%%%%%%%%%%%%%%%%%%%%%%%%%%%

The atmosphere adds  several complications to our analysis  that we do
not  have to deal  with for  $HST$.  The  obvious complication  is the
seeing, which limits our ability  to resolve stars and measure precise
positions, and  makes the PSF change  from image to image.   This is a
random effect,  but there are  also systematic effects such  as: image
motion   due  to  isoplanatic   patches  and   differential  chromatic
refraction.

We must postpone to a future paper of this series a more comprehensive
study  of the  atmospheric effects,  when a  better data-base  will be
available\footnote{At $WFI@2.2m$  60 hours have  already been approved
to us for 2006.}.  The images here are too few, and taken at a limited
range of airmasses.   In this section, we will  present results from a
few  tests, to  quantify biases  and give,  a posteriori,  some simple
corrections.

%%%%%%%%%%%
\subsection{Image Motion}
%%%%%%%%%%%

One of the atmospheric effects we  should be aware of is image motion.
Atmospheric  turbulence introduces  perturbations at  different scales
for different  integration times.  If  we take very  short ``speckle''
images of a  small field, we can get better resolution  at the cost of
less  photons.   This  is  because  all  of  the  photons  go  through
essentially the same patch of  atmosphere and are all shifted the same
way  by the  atmosphere.  If  longer exposures  are taken,  the moving
packets of air will cause the small field of view to shift up and down
together, blurring out  the image.  As the exposure  time goes up, the
amplitude  of  the  coherent  shifts  goes down  while  the  scale  of
coherence  also   goes  down.   Fig.~6  in  Platais   et  al.\  (2002)
illustrates the effect of the atmosphere directly.

Lindegren (1980) provides  a simple formulation to give  us an idea of
what we can expect to see from the atmosphere in each coordinate:\
$$
\sigma_T \ [arcsec]=0''.8 \ \mathcal{R}_{\rm [rad]}^{0.25} \ T_{\rm [s]}^{-0.5},
$$
where $T  \gg 300 ~ \mathcal{R}$  is the integration  time in seconds.
For the $WFI@2.2m$ field, typical $\mathcal{R}_{\rm max} \simeq 0.005$
rad (from  center to edge of  our field), so for  the NGC~2477 images,
with  exposure times  of 900s,  the  maximum global  effect we  should
expect is 6.9  mas.  This is in good agreement  with what was obtained
in  Section\ \ref{dein},  and shown  in Fig.\  \ref{rms1}  ({\it lower
panel}).  Zacharias (1996) found  the global effects of the atmosphere
on  differential astrometry  to  be smaller  than  the predictions  of
Lindegren  (1980), so our  global residuals  are likely  indicative of
small distortion  errors (as discussed  in Section 5), in  addition to
the   atmospheric  effects.    At   any  rate,   our   use  of   local
transformations  will  minimize  our  sensitivity to  both  distortion
errors and larger-scale atmospheric effects.

%%%%%%%%%%%
\subsection{Differential Chromatic Refraction}
%%%%%%%%%%%

The effects of  image motion above will average out  if we take enough
long  exposures, but  differential  chromatic refraction  effect is  a
systematic effect that will not average out with more observations.

Differential  chromatic refraction  (DCR) causes  a shift  between the
centroid of the  blue photons and that of the  red photons.  This will
cause blue stars  to have more of a shift towards  the zenith than the
red stars will have.  When we observe through a filter, this effect is
lessened, since all of the photons have about the same wavelength, but
the details of the spectral  distribution through the filter can still
affect the centroid of a star's position.  This is an effect that will
not  go away with  linear transformations,  so we  must be  careful to
calibrate and remove it.

We note  that the  linearity of  CCDs makes it  easier to  observe and
remove this  effect.  With the old technology  of photographic plates,
the  shape  of  a star  and  its  apparent  centroid depended  on  its
brightness, due to non-linearity effects.  Since colors and magnitudes
are  typically strongly  correlated  in  a CMD,  it  was difficult  to
independently  remove the  atmospheric and  photographic  effect.  The
linearity of CCDs allows us to remove much of this degeneracy.

The  best  way  to  calibrate  the  DCR effect  is  to  take  multiple
observations at a  variety of zenith angles, as  described in Monet et
al. (1992). 
(See also Stone et al. (2003), and the references therein.)
Since  the data  set examined  here was  not optimized  for calibrated
astrometry, we cannot  solve for and remove the  DCR effect.  However,
we can  still examine it.  If  there was a difference  between the DCR
effect at one epoch compared to  another, then we would expect the DCR
effect to generate  an apparent proper motion for  blue stars relative
to red stars.

In Figures 16 and 17, we show the proper motion as a function of color
for stars  on the horizontal and  giant branches.  Both  sets of stars
should be moving  with the cluster and have no  motion relative to the
MS  stars.   Instead,  we  see  that  in  both  clusters  there  is  a
color-related displacement of $\sim$0.05  pixel ($\sim$12 mas) for M4,
and a displacement of $\sim$0.02 pixel ($\sim$5 mas) for NGC~6397.

\begin{figure}
   \centering
   \includegraphics[width=8.5cm]{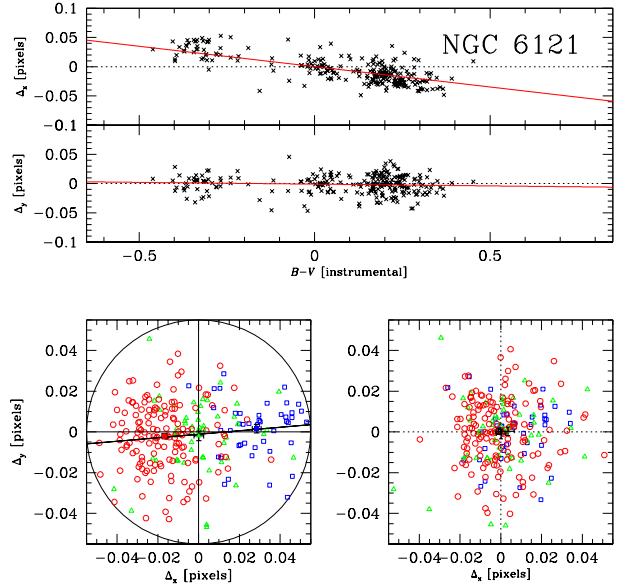}
      \caption{
%	%
	{\em Bottom-Left:} In  this panel we zoom in  the vector point
	diagram of Fig.\ \ref{n6121_II}  for stars in the instrumental
	$B$-magnitude  between  $-16$ and  $-14$  for NGC~6121  before
	applying DCR corrections. RGB stars are marked as circles, red
	HB  stars as  triangles, and  blue HB  with squares.  Also the
	average for  each group (with  error bars) are shown,  using a
	filled symbols.
%	%	
	{\em Top: } The distribution of displacements as a function of
	the color is modeled with a linear fit.
%	%	
	{\em   Bottom-Right:}   The   same  distribution   after   the
	corrections obtained from the linear fit.
%	%
	}
         \label{n6121_DCR}
   \end{figure}

\begin{figure}
   \centering
   \includegraphics[width=8.5cm]{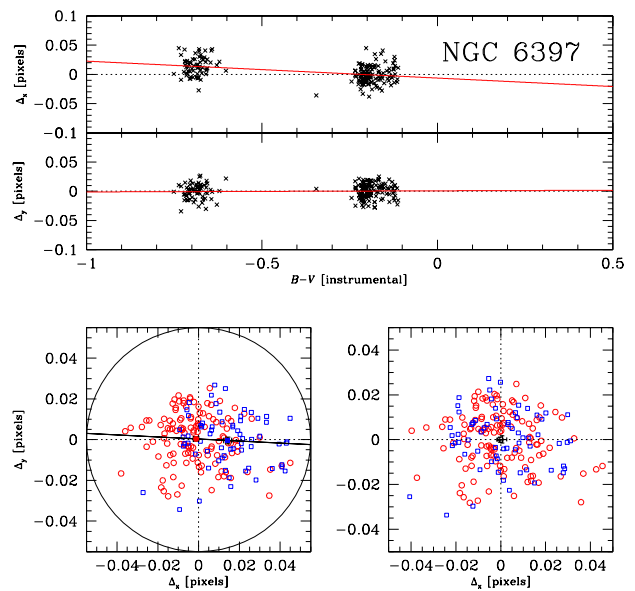}
      \caption{
%	%
	As in Fig.\ \ref{n6121_DCR},  but for NGC~6397.  This time the
	stars are  the one of the  instrumental $B$-magnitude interval
	$-16$-$-14$ of  Fig.\ \ref{n6397_II}.  NGC~6397 has no  red HB
	stars.
% 	%
	}
         \label{n6397_DCR}
   \end{figure}

We made  linear fits to the  distribution of points shown  in Figs. 16
and 17 ({\it upper panels}), and removed the DCR contribution from the
proper  motions.  The  final, corrected  proper motions  are  shown in
Figs.   \ref{n6121_DCR},  and   \ref{n6397_DCR}  ({\it   bottom  right
panels}).  In  properly planned  observations, this correction  can be
made for each image, as a function of the airmass of the observations.
The correction  can also be done all  within one epoch, so  that we do
not need to assume anything about the proper motions.

%%%%%%%%%%%%%%%%%%%%%%%%%%%%%%%%%%%%%%%%%%%%%%%%%%%%%%%%%%%%%%%%%
%
%
% FIGX --- 
%______________________________________________________________
%
%_________________________________________________________________
%

%%%%%%%%%%%%%%%%%%%%%%%%%%%%%%%%%%%%%%%%%%%%%%%%%%%%%%%%%%%%%%%%%
%
%
% FIGX --- 
%______________________________________________________________
%
%_________________________________________________________________
%

%%%%%%%%%%%%%%%%%%%%%%%%%%%%%%%%%%%%%%%%%%%%%%%%%%%%%%%%%%%%%%%%%
%
\section{Conclusion and future applications}
\label{CON}
%
%%%%%%%%%%%%%%%%%%%%%%%%%%%%%%%%%%%%%%%%%%%%%%%%%%%%%%%%%%%%%%%%%

In this paper, we have described how the software originally developed
by Anderson and King (2000) for high precision relative astrometry and
photometry  on  WFPC2   and  ACS  $HST$  data  has   been  adapted  to
ground-based, wide  field images from the  WFI camera at  the ESO 2.2m
telescope. We  have also obtained  a first approximation  solution for
the $WFI@2.2m$ geometric  distortion, and shown that it  is not stable
over time.  Therefore, to get precise relative proper motions, we need
to follow a local-transformation approach, as described in Section 7.

As proof of concept, we have  applied this new technique on two epochs
of data for the two  closest Galactic globular clusters: NGC 6121 (M4)
and NGC  6397.  The  results, though based  on data not  optimized for
high-precision  astrometric measurements,  are more  than encouraging.
We have shown  that, under average seeing conditions  ($\sim 1''$) the
astrometric precision is of 7 mas in each coordinate, for well exposed
stars in a single image, i.e.   only $\sim$3 (6) times worse than what
we are able to obtain with $HST$ using the WFPC2 (WFC/ACS) (which more
or  less represents  the {\em  ``the state  of the  art''}  in imaging
astrometry).

For both clusters, with three-year temporal baseline we have been able
to obtain proper-motion measurements  that are precise enough to allow
a separation between field and cluster stars.  We expect to be able to
measure  the internal  proper motions  with a  precision  adequate for
stellar dynamics studies with a  10-year baseline and a good number of
images at each epoch.

The  extension  of the  AK2000  software  to  ground-based data  makes
possible a great  number of new projects for the  study of the stellar
population  in Galactic  open and  globular clusters  and  their tidal
tails, in the Galactic fields,  and in nearby galaxies, and represents
an  important  complement  to   what  is  presently  done  with  $HST$
images. Ground-based  facilities are more abundant,  allow coverage of
much larger fields,  and are cheaper and easier  to access than $HST$.
$HST$ data  are still of  fundamental importance for the  most crowded
regions of star clusters and nearby galaxies.

The forthcoming even-wider-field  facilities (e.g. OMEGACAM@VST, VISTA,
etc.),  and the  increased  time baseline  (when  we include  archival
first-epoch  data  taken in  the  mid-90's)  will  allow to  furtherly
exploit the technique here described.

This  technique may  also  be  promising in  view  of the  soon-coming
multi-conjugated  adaptive   optics,  and  non-classical  narrow-field
astrometric   corrections  for  telescopes   larger  than   10  meters
(Lazorenko \& Lazorenko, 2004, Lazorenko 2005).

Finally,  it is  worth mentioning  that  most of  the astrometric  and
proper-motion  measurements  on  $HST$  and  ground-based  images  are
complementary to the  data expected from SIM and  GAIA.  First of all,
GAIA is  several years away  in the future.  Realistically,  the final
catalog will  be released not before  the end of the  second decade of
this century, at best. Also  the catalog will be limited to magnitudes
brighter  than $V\sim19$,  and,  most importantly,  to  stars in  less
crowded regions.

Over the next few years, we will continue to improve these techniques.
Specifically, in the  next months, we will have  $WFI@2.2m$ data for 5
open clusters (60 hours of  observing time already scheduled). The new
data,  properly   dithered  in  order  to   optimize  the  astrometric
measurements,  will  allow   further  improvement  of  the  distortion
solution, and more study of atmospheric effects on our astrometry.

%__________________________________________________ 
% 

\begin{acknowledgements}
We thank the referee Dr.  Norbert Zacharias for careful reading of the
manuscript and for many useful comments.
GP, and  RSY, acknowledge  the support of  the MIUR under  the program
PRIN2003.  JA acknowledges the support of NASA/HST grant GO-10252.
\end{acknowledgements}

%______________________________________________________________
%

\end{document}